\documentclass[draftcls, onecolumn, 11pt]{IEEEtran}
\usepackage{cite,color,graphicx,url,amssymb,amsthm,threeparttable,multirow,algorithm,algorithmic}
\usepackage[tight,footnotesize]{subfigure}
\usepackage[cmex10]{amsmath}
\IEEEoverridecommandlockouts

\newcommand{\FIGSCALE}{0.29}
\newcommand{\cH}{\mathcal{H}}
\newcommand{\cT}{\mathcal{T}}
\newcommand{\cW}{\mathcal{W}}

\newcommand{\m}[1]{\mathcal{#1}}
\newcommand{\bs}[1]{\boldsymbol{#1}}
\newcommand{\mbf}[1]{\mathbf{#1}}
\newcommand{\Dft}[2]{\mathbf{#1}_{#2}\left(e^{j\omega T}\right)}
\newcommand{\dft}[2]{{#1}_{#2}\left(e^{j\omega T}\right)}
\newcommand{\defn}{\stackrel{}{=}}

\theoremstyle{plain}

\newtheorem{theorem}{Theorem}
\newtheorem{corollary}{Corollary}
\newtheorem{proposition}{Proposition}

\newtheorem{condition}{Condition}
\theoremstyle{definition}

\theoremstyle{remark}
\newtheorem{remark}{Remark}

\begin{document}

% paper title
\title{Identification of Parametric Underspread Linear Systems and Super-Resolution Radar}

% authors
\author{Waheed~U.~Bajwa, \IEEEmembership{Member,~IEEE}, Kfir~Gedalyahu, and Yonina~C.~Eldar, \IEEEmembership{Senior~Member,~IEEE}%
% thanks
\thanks{W.U. Bajwa is with the Department of Electrical and Computer Engineering, Duke University, Durham, NC 27707 USA (Email: {\tt w.bajwa@duke.edu}). K. Gedalyahu and Y.C. Eldar are with the Department of Electrical Engineering, Technion---Israel Institute of Technology, Haifa 32000, Israel (Phone: +972-4-8293256, Fax: +972-4-8295757, E-mails: {\tt \{kfirge@techunix,yonina@ee\}.technion.ac.il}. Y.C. Eldar is currently also a Visiting Professor at Stanford University, Stanford, CA 94305 USA.}
}

% running headings
%\markboth{}{}

% make the title area
\maketitle

%***************
\begin{abstract}
Identification of time-varying linear systems, which introduce both time-shifts (delays) and frequency-shifts (Doppler-shifts), is a central task in many engineering applications. This paper studies the problem of identification of underspread linear systems (ULSs), whose responses lie within a unit-area region in the delay--Doppler space, by probing them with a known input signal. It is shown that sufficiently-underspread parametric linear systems, described by a finite set of delays and Doppler-shifts, are identifiable from a single observation as long as the time--bandwidth product of the input signal is proportional to the square of the total number of delay--Doppler pairs in the system. In addition, an algorithm is developed that enables identification of parametric ULSs from an input train of pulses in polynomial time by exploiting recent results on sub-Nyquist sampling for time delay estimation and classical results on recovery of frequencies from a sum of complex exponentials. Finally, application of these results to super-resolution target detection using radar is discussed. Specifically, it is shown that the proposed procedure allows to distinguish between multiple targets with very close proximity in the delay--Doppler space, resulting in a resolution that substantially exceeds that of standard matched-filtering based techniques without introducing leakage effects inherent in recently proposed compressed sensing-based radar methods.
\end{abstract}

%***************
\begin{IEEEkeywords}
Compressed sensing, Delay--Doppler estimation, rotational invariance techniques, super-resolution radar, system identification, time-varying linear systems
\end{IEEEkeywords}

%***************
\section{Introduction}
Physical systems arising in a number of application areas can often be described as linear and time varying \cite{proakis:01,skolnik:01}. Identification of such systems may help improve overall performance, e.g., the bit-error rate in communications \cite{proakis:01}, or constitute an integral part of the overall system operation, e.g., target detection using radar or active sonar \cite{skolnik:01}.

\begin{figure*}
\begin{center}
\includegraphics[scale=0.5]{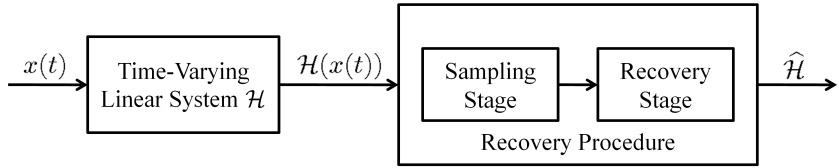}
\caption{\label{fig:sys_id}Schematic representation of identification of a time-varying linear system $\cH$ by probing it with a known input signal. Characterization of an identification scheme involves specification of the input probe, $x(t)$, and the accompanying sampling and recovery stages.}
\end{center}
\end{figure*}

Mathematically, identification of a given time-varying linear system $\cH$ involves probing it with a known input signal $x(t)$ and identifying $\cH$ by analyzing the single system output $\cH(x(t))$ \cite{kozek:sjma05}, as illustrated in Fig.~\ref{fig:sys_id}. Unlike time-invariant linear systems, however, a single observation of a time-varying linear system does not lead to a unique solution unless additional constraints on the system response are imposed. This is due to the fact that such systems introduce both time-shifts (\emph{delays}) and frequency-shifts (\emph{Doppler-shifts}) to the input signal. It is now a well-established fact in the literature that a time-varying linear system $\cH$ can only be identified from a single observation if $\cH(\delta(t))$ is known to lie within a region $\m{R}$ of the delay--Doppler space such that ${\tt area}(\m{R}) < 1$ \cite{kailath:tit62,bello:tit69,kozek:sjma05,pfander:tit06}. Identifiable time-varying linear systems are termed \emph{underspread}, as opposed to nonidentifiable \emph{overspread} linear systems, which satisfy ${\tt area}(\m{R}) > 1$ \cite{kozek:sjma05,pfander:tit06}.\footnote{It is still an open research question as to whether \emph{critically-spread} linear systems, which correspond to ${\tt area}(\m{R}) = 1$, are identifiable or nonidentifiable \cite{pfander:tit06}; see \cite{kozek:sjma05} for a partial answer to this question for the case when $\m{R}$ is a rectangular region.}

In this paper, we study the problem of identification of underspread linear systems (ULSs) whose responses can be described by a finite set of delays and Doppler-shifts. That is,
\begin{align}
\label{eqn:LTV_sys1}
 \cH(x(t)) = \sum_{k=1}^K \alpha_k x(t - \tau_k) e^{j 2 \pi \nu_k t}
\end{align}
where $(\tau_k, \nu_k)$ denotes a delay--Doppler pair and $\alpha_k \in \mathbb{C}$ is the complex attenuation factor associated with $(\tau_k, \nu_k)$. Unlike most of the existing work in the literature, however, our goal in this paper is to explicitly characterize conditions on the bandwidth and temporal support of the input signal that ensure identification of such ULSs from single observations. The importance of this goal can be best put into perspective by realizing that ULSs of the form \eqref{eqn:LTV_sys1} tend to arise frequently in many applications. Consider, for example, a single-antenna transmitter communicating wirelessly with a single-antenna receiver in a mobile environment. Over a small-enough time interval, the channel between the transmitter and receiver is of the form \eqref{eqn:LTV_sys1} with each triplet $(\tau_k, \nu_k, \alpha_k)$ corresponding to a distinct physical path between the transmitter and receiver \cite{bajwa:proc10}. Identification of $\cH$ can enable one to establish a relatively error-free communication link between the transmitter and receiver. But wireless systems also need to identify channels using signals that have as small \emph{time--bandwidth product} as possible so that they can allocate the rest of the temporal degrees of freedom to communicating data \cite{slepian:pieee76,bajwa:proc10}.

Similarly, in the case of target detection using radar or active sonar, the (noiseless, clutter-free) received signal is of the form \eqref{eqn:LTV_sys1} with each triplet $(\tau_k, \nu_k, \alpha_k)$ corresponding to an echo of the transmitted signal from a distinct target in the delay--Doppler space \cite{skolnik:01}. Identification of $\cH$ in this case enables one to accurately obtain the radial position and velocity of the targets. Radar systems also strive to operate with signals (waveforms) that have as small temporal support and bandwidth as possible. This is because the temporal support of the radar waveform is directly tied to the time it takes to identify all the targets while the bandwidth of the waveform---among other technical considerations---is tied to the sampling rate of the radar receiver \cite{skolnik:01}.

Given the ubiquity of time-varying linear systems in engineering applications, there exists considerable amount of existing literature that studies identification of such systems in an abstract setting. Kailath was the first to recognize that the identifiability of a time-varying linear system $\cH$ from a single observation is directly tied to the area of the region $\m{R}$ that contains $\cH(\delta(t))$ \cite{kailath:tit62}. Kailath's seminal work in \cite{kailath:tit62} laid the foundations for the future works of Bello \cite{bello:tit69}, Kozek and Pfander \cite{kozek:sjma05}, and Pfander and Walnut \cite{pfander:tit06}, which establish the nonidentifiability of overspread linear systems and provide constructive proofs for the identifiability of arbitrary ULSs. However, none of these results shed any light on the bandwidth and temporal support of the input signal needed to ensure identification of ULSs of the form \eqref{eqn:LTV_sys1}. On the contrary, the constructive proofs provided in \cite{kailath:tit62, bello:tit69, kozek:sjma05, pfander:tit06} require use of input signals with infinite bandwidth and temporal support.

In contrast, to the best of our knowledge, this is the first paper to develop a theory for identification of ULSs of the form \eqref{eqn:LTV_sys1}, henceforth referred to as \emph{parametric} ULSs, that parallels that of \cite{kailath:tit62, bello:tit69, kozek:sjma05, pfander:tit06} for identification of arbitrary ULSs. One of the main contributions of this paper is that we establish using a constructive proof that sufficiently-underspread parametric linear systems are identifiable as long as the time--bandwidth product of the input signal is proportional to the square of the total number of delay--Doppler pairs in the system. Equally importantly, as part of our constructive proof, we concretely specify the nature of the input signal (a finite train of pulses) and the structure of a corresponding polynomial-time (in the number of delay--Doppler pairs) recovery procedure that enable identification of parametric ULSs. These ideas are also immediately applicable to super-resolution target detection using radar and we show in the latter part of the paper that our approach indeed results in a resolution that substantially exceeds that of standard matched-filtering based techniques without introducing leakage effects inherent in recently proposed compressed sensing-based radar methods \cite{herman:tsp09}.

The key developments in the paper leverage recent results on sub-Nyquist sampling for time-delay estimation \cite{LRTDE} and classical results on direction-of-arrival (DOA) estimation \cite{KT, 1164935, hua1990matrix, stoica1997}. Unlike the traditional DOA estimation literature, however, we do not assume that the system output is observed by an array of antennas. Additionally, in contrast to \cite{LRTDE}, our goal here is not a reduction in the sampling rate; rather, we are interested in characterizing the minimum temporal degrees of freedom of the input signal needed to ensure identification of a parametric ULS $\cH$. The connection to sub-Nyquist sampling can be understood by noting that the sub-Nyquist sampling results of \cite{LRTDE} enable recovery of the delays associated with $\cH$ using a small-bandwidth input signal. Further, the ``train-of-pulses'' nature of the input signal combined with results on recovery of frequencies from a sum of complex exponentials \cite{stoica1997} allow recovery of the Doppler-shifts and attenuation factors using an input signal of small temporal support.

Several works in the past have considered identification of specialized versions of parametric ULSs. Specifically, \cite{kay:icassp97, wong:globecom06, bajwa:allerton08, herman:tsp09, tan:tsp09} treat parametric ULSs whose delays and Doppler-shifts lie on a quantized grid in the delay--Doppler space. On the other hand, \cite{jakobsson:tsp98} considers the case in which there are no more than two Doppler-shifts associated with the same delay. The proposed recovery procedures in \cite{jakobsson:tsp98} also have exponential complexity, since they require exhaustive searches in a $K$-dimensional space, and stable initializations of these procedures stipulate that the system output be observed by an $M$-element antenna array with $M \gtrapprox K$.

While the insights of \cite{kay:icassp97, wong:globecom06, bajwa:allerton08, herman:tsp09, tan:tsp09} can be extended to arbitrary parametric ULSs by taking infinitesimally-fine quantization of the delay--Doppler space, this will require input signals with infinite bandwidth and temporal support. In contrast, our ability to avoid quantization of the delay--Doppler space is due to the fact that we treat the system-identification problem directly in the analog domain. This follows the philosophy in much of the recent work in analog compressed sensing, termed Xampling, which provides a framework for incorporating and exploiting structure in analog signals without the need for quantization \cite{MEDS09,MEE10,ME09c,ME09d,E09,EM09a}. This is in particular the key enabling factor that helps us avoid the catastrophic implications of the leakage effects in the context of radar target detection.

Before concluding this discussion, we note that responses of arbitrary ULSs can always be represented as \eqref{eqn:LTV_sys1} under the limit $K \rightarrow \infty$. Therefore, the main result of this paper can also be construed as an alternate constructive proof of the identifiability of sufficiently-underspread linear systems. Nevertheless, just like \cite{kailath:tit62, bello:tit69, kozek:sjma05, pfander:tit06}, this interpretation of the presented results again seem to suggest that identification of arbitrary ULSs requires use of input signals with infinite bandwidth and temporal support.

The rest of this paper is organized as follows. In Section~\ref{sec:prob}, we formalize the problem of identification of parametric ULSs along with the accompanying assumptions. In Section~\ref{sec:id_sch}, we propose a polynomial-time recovery procedure used for the identification of parametric ULSs, while Section~\ref{sec:suff_cond} specifies the conditions on the input signal needed to guarantee unique identification using the proposed procedure. We compare the results of this paper to some of the related literature on identification of parametric ULSs in Section~\ref{sec:disc} and discuss an application of our results to super-resolution target detection using radar in Section~\ref{sec:radar}. Finally, we present results of some numerical experiments in Section~\ref{sec:num_res}.

We make use of the following notational convention throughout this paper. Vectors and matrices are denoted by bold-faced lowercase and bold-faced uppercase letters, respectively. The $n$th element of a vector $\mathbf{a}$ is written as $\mathbf{a}_{n}$, and the $(i,j)$th element of a matrix $\mathbf{A}$ is denoted by $\mathbf{A}_{ij}$. Superscripts $\left(\cdot\right)^{*}$, $\left(\cdot\right)^{T}$ and $\left(\cdot\right)^{H}$ represent conjugation, transposition, and conjugate transposition, respectively. In addition, the Fourier transform of a continuous-time signal $x\left(t\right)\in L_{2}(\mathbb{C})$ is defined by $X\left(\omega\right) \defn \int_{-\infty}^{\infty}x\left(t\right)e^{-j\omega t}dt$, while $\left\langle x\left(t\right),y\left(t\right)\right\rangle = \int_{-\infty}^{\infty}x\left(t\right)y^{*}(t)dt$ denotes the inner product between two continuous-time signals in $L_{2}(\mathbb{C})$. Similarly, the discrete-time Fourier transform of a sequence $a\left[n\right]\in\ell_{2}(\mathbb{C})$ is defined by $A\left(e^{j\omega T}\right)=\sum_{n\in\mathbb{Z}}a\left[n\right]e^{-j\omega nT}$ and is periodic in $\omega$ with period $2\pi/T$. Finally, we use $\mathbf{A}^{\dagger}$ to write the Moore--Penrose pseudoinverse of a matrix $\mathbf{A}$.

%%%%%%%%%%%%%%%%%%%%%%%%%%%%%%%%%%%%%%%%%%%%%%%%%%%%%%%%%%%%%
\section{Problem Formulation and Main Results}\label{sec:prob}
In this section, we formalize the problem of identification of a parametric ULS $\cH$ whose response is described by a total of $K$ \emph{arbitrary} delay--Doppler-shifts of the input signal. The task of identification of $\cH$ essentially requires specifying two distinct but highly intertwined steps. First, we need to specify the conditions on the bandwidth and temporal support of the input signal $x(t)$ that ensure identification of $\cH$ from a single observation. Second, we need to provide a polynomial-time recovery procedure that takes as input $\cH(x(t))$ and provides an estimate $\widehat\cH$ of the system response by exploiting the properties of $x(t)$ specified in the first step. We begin by detailing our system model and the accompanying assumptions.

In \eqref{eqn:LTV_sys1}, some of the delays, $\tau_k$, might be repeated. Expressing $\cH$ in terms of $K_\tau \leq K$ \emph{distinct} delays in this case leads to
\begin{align}
\label{eqn:LTV_sys}
	\cH(x(t)) = \sum_{i=1}^{K_\tau} \sum_{j=1}^{K_{\nu,i}} \alpha_{ij} x(t - \tau_i) e^{j 2 \pi \nu_{ij} t}
\end{align}
where $\nu_{ij}$ denotes the $j$th Doppler-shift associated with the $i$th distinct delay $\tau_i$,  $\alpha_{ij} \in \mathbb{C}$ denotes the attenuation factor associated with the delay--Doppler pair $(\tau_i, \nu_{ij})$, and $K \defn \sum_{i=1}^{K_\tau} K_{\nu,i}$. We choose to express $\cH(x(t))$ as in \eqref{eqn:LTV_sys} so as to facilitate the forthcoming analysis. Throughout the rest of the paper, we use $\bs{\tau} \defn \{ \tau_{i}, \ i=1,\dots,K_\tau\}$ to denote the set of $K_\tau$ distinct delays associated with $\cH$. The first main assumption that we make concerns the footprint of $\cH$ in the delay--Doppler space:
\begin{enumerate}
 \item[\textbf{[A1]}] The response $\cH(\delta(t))$ of $\cH$ lies within a rectangular region of the delay--Doppler space; in other words, $(\tau_i, \nu_{ij}) \in [0, \tau_{max}] \times [-\nu_{max}/2, \nu_{max}/2]$. This is indeed the case in many engineering applications (see, e.g., \cite{proakis:01,skolnik:01}). The parameters $\tau_{max}$ and $\nu_{max}$ are termed in the parlance of linear systems as the \emph{delay spread} and the \emph{Doppler spread} of the system, respectively.
\end{enumerate}

Next, we use $\cT$ and $\cW$ to denote the temporal support and the two-sided bandwidth of the known input signal $x(t)$ used to probe $\cH$, respectively. We propose using input signals that correspond to a finite train of pulses:
\begin{align}
\label{eqn:input_probe}
	x(t) = \sum_{n=0}^{N-1} x_n g(t - nT), \ 0 \leq t \leq \cT
\end{align}
where $g(t)$ is a prototype pulse of bandwidth $\cW$ that is (essentially) temporally supported on $[0,T]$ and is assumed to have unit energy $(\int |g(t)|^2 dt = 1)$, and $\{x_n \in \mathbb{C}\}$ is an $N$-length probing sequence. The parameter $N$ is proportional to the time--bandwidth product of $x(t)$, which roughly defines the number of temporal degrees of freedom available for estimating $\cH$ \cite{slepian:pieee76}: $N = \cT/T \propto \cT\cW$.\footnote{Recall that the temporal support and the bandwidth of an arbitrary pulse $g(t)$ are related to each other as $\cW \propto 1/T$.} The final two assumptions that we make concern the relationship between the delay spread $\tau_{max}$  and the Doppler spread $\nu_{max}$ of $\cH$, and the temporal support $T$ and bandwidth $\cW$ of $g(t)$:
\begin{enumerate}
 \item[\textbf{[A2]}] The delay spread of $\cH$ is strictly smaller than the temporal support of $g(t)$: $\tau_{max} < T$, and
 \item[\textbf{[A3]}] The Doppler spread of $\cH$ is much smaller than the bandwidth of $g(t)$: $\nu_{max} \ll \cW$.
\end{enumerate}

Note that, since $\cW \propto 1/T$, \textbf{[A3]} equivalently imposes that $\nu_{max} T \ll 1$. This assumption states that the temporal scale of variations in $\cH$ is large relative to the temporal scale of variations in $x(t)$. It is worth pointing out that linear systems that are sufficiently underspread in the sense that $\tau_{max} \nu_{max} \ll 1$ can always be made to satisfy \textbf{[A2]} and \textbf{[A3]} for any given budget of the time--bandwidth product.
\begin{remark}
In order to elaborate on the validity of \textbf{[A2]} and \textbf{[A3]}, note that there exist many communication applications where underlying linear systems tend to be highly underspread \cite[\S~14.2]{proakis:01}. Similarly, \textbf{[A2]} and \textbf{[A3]} in the context of radar target detection simply mean that the targets are not too far away from the radar and that their velocities are not too high. Consider, for example, an $L$-band radar (operating frequency of $1.3$ GHz) that transmits a pulse $g(t)$ of bandwidth $\cW = 10$ MHz after every $T = 50~\mu$s. Then both \textbf{[A2]} and \textbf{[A3]} are satisfied when the distance between the radar and any target is at most $7.5$ km $(\tau_{max} \approx 50~\mu\text{s})$ and the radial velocity of any target is at most $185$ km/h $(\nu_{max} \approx 445 \text{ Hz})$ \cite{skolnik:01}.
\end{remark}

The following theorem summarizes our key result concerning identification of parametric ULSs.
\begin{theorem}[Identification of Parametric Underspread Linear Systems]
\label{thm:pf_results}
Suppose that $\cH$ is a parametric ULS that is completely described by a total of $K = \sum_{i=1}^{K_\tau} K_{\nu,i}$ triplets $(\tau_i, \nu_{ij}, \alpha_{ij})$. Then, irrespective of the distribution of $\{(\tau_i, \nu_{ij})\}$ within the delay--Doppler space, $\cH$ can be identified in polynomial-time from a single observation $\cH(x(t))$ as long as it satisfies \textbf{[A1]}--\textbf{[A3]}, the probing sequence $\{x_n\}$ remains bounded away from zero in the sense that $|x_n| > 0$ for every $n=0,\dots,N-1$, and the time--bandwidth product of the known input signal $x(t)$ satisfies the condition
\begin{align}
\label{eqnthm:TW}
 \mathcal{TW} \geq 8 \pi K_\tau K_{\nu,max}
\end{align}
where $K_{\nu,max} \defn \max_i K_{\nu,i}$ is the maximum number of Doppler-shifts associated with any one of the distinct delays. In addition, the time--bandwidth product of $x(t)$ is guaranteed to satisfy \eqref{eqnthm:TW} as long as $\mathcal{TW} \geq  2 \pi (K+1)^2$.
\end{theorem}
\noindent The rest of this paper is devoted to providing a proof of Theorem~\ref{thm:pf_results}. In terms of a general roadmap for the proof, we first exploit the sub-Nyquist sampling results of \cite{LRTDE} to argue that $x(t)$ with small bandwidth suffices to recover the delays associated with $\cH$. We then exploit the ``train-of-pulses'' structure of $x(t)$ and classical results on recovery of frequencies from a sum of complex exponentials \cite{stoica1997} to argue that $x(t)$ with small temporal support suffices to recover the Doppler-shifts and attenuation factors associated with $\cH$. The statement of Theorem~\ref{thm:pf_results} will then follow by a simple combination of the two claims concerning the bandwidth and temporal support of $x(t)$. We will make use of \eqref{eqn:LTV_sys} and \eqref{eqn:input_probe} in the following to describe:
\begin{enumerate}
 \item[\textbf{[1]}] The polynomial-time recovery procedure used for the identification of $\cH$ (cf.~Section~\ref{sec:id_sch}), and
 \item[\textbf{[2]}] The accompanying conditions on $x(t)$ needed to guarantee identification of $\cH$ (cf.~Section~\ref{sec:suff_cond}).
\end{enumerate}

%%%%%%%%%%%%%%%%%%%%%%%%%%%%%%%%%%%%%%%%%%%%%%%%%%%%%%%%%%%%%
\section{Polynomial-Time Identification of ULSs}\label{sec:id_sch}
\begin{figure*}
\begin{center}
\includegraphics[scale=0.375]{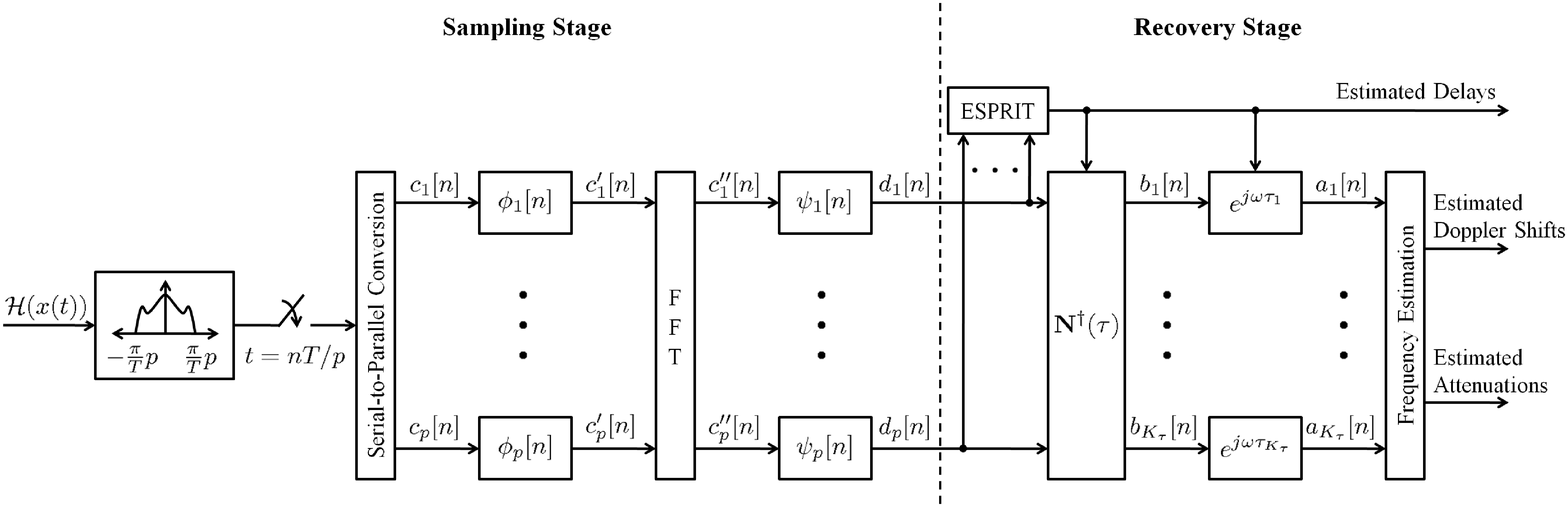}
\caption{\label{fig:samp_rec_scheme}Schematic representation of the polynomial-time recovery procedure for identification of parametric underspread linear systems from single observations.}
\end{center}
\end{figure*}
In this section, we characterize the polynomial-time recovery procedure used for identification of ULSs of the form \eqref{eqn:LTV_sys}. In order to facilitate understanding of the proposed algorithm, shown in Fig.~\ref{fig:samp_rec_scheme}, we conceptually partition the method into two stages: sampling and recovery. The rest of this section is devoted to describing these two steps in detail. Before proceeding further, however, it is instructive to first make use of \eqref{eqn:LTV_sys} and \eqref{eqn:input_probe} and rewrite the output of $\cH$ as
\begin{align}
\nonumber
	\cH(x(t)) &= \sum_{i=1}^{K_\tau} \sum_{j=1}^{K_{\nu,i}} \sum_{n=0}^{N-1} \alpha_{ij} x_n e^{j 2 \pi \nu_{ij} t} g(t - \tau_i - nT)\\
\nonumber
	  &\stackrel{(a)}{\approx} \sum_{i=1}^{K_\tau} \sum_{j=1}^{K_{\nu,i}} \sum_{n=0}^{N-1} \alpha_{ij} x_n e^{j 2 \pi \nu_{ij} nT} g(t - \tau_i - nT)\\
\label{eq:received_sig}
	  &= \sum_{i=1}^{K_\tau} \sum_{n=0}^{N-1} a_i[n] g(t - \tau_i - nT)
\end{align}
where $(a)$ follows from the assumption $\nu_{max} T \ll 1$, which implies that $e^{j 2 \pi \nu_{ij} t} \approx e^{j 2 \pi \nu_{ij} nT}$ for all $t \in [(n-1)T, n T)$, and the sequences $\{a_i[n]\}, i=1,\dots,K_\tau$, are defined as
\begin{align}
\label{eq:sequences}
	a_i[n] = \sum_{j=1}^{K_{\nu,i}} \alpha_{ij} x_n e^{j 2 \pi \nu_{ij} nT}, \ n=0,\dots,N-1.
\end{align}

\subsection{The Sampling Stage}
We leverage the ideas of \cite{LRTDE} on time-delay estimation from sub-Nyquist samples to describe the sampling stage of our recovery procedure. While the primary objective in \cite{LRTDE} is time-delay estimation from low-rate samples, the development here is carried out with an eye towards identification of parametric ULSs regardless of the distribution of system parameters within the delay--Doppler space---the so-called \emph{super-resolution identification}. In \cite{LRTDE}, a general multi-channel sub-Nyquist sampling scheme was introduced for the purpose of recovering a set of unknown delays from signals of the form \eqref{eq:received_sig}. Here, we focus on one special case of that scheme, which consists of a low-pass filter (LPF) followed by a uniform sampler. This architecture may be preferable from an implementation viewpoint since it requires only one sampling channel, thereby simplifying analog front-end of the sampling hardware. The LPF, besides being required by the sampling stage, also serves as the front-end of the system-identification hardware and rejects noise and interference outside the working spectral band.

Our sampling stage first passes the system output $y(t) \defn \cH(x(t))$ through a LPF whose impulse response is given by $s^*(-t)$ and then uniformly samples the LPF output at times $\big\{t=nT/p\big\}$. We assume that the frequency response, $S^*(\omega)$, of the LPF is contained in the spectral band $\m{F}$, defined as
\begin{align}
\m{F}=\left[-\frac{\pi}{T}p,\frac{\pi}{T}p\right],
\end{align}
and is zero for frequencies $\omega \notin \m{F}$. Here, the parameter $p$ is assumed to be even and satisfies the condition $p \geq 2K_\tau$; exact requirements on $p$ to ensure identification of $\cH$ will be given in Section~\ref{sec:suff_cond}. In order to relate the sampled output of the LPF with the multi-channel sampling formulation of \cite{LRTDE}, we define $p$ sampling (sub)sequences $\big\{c_{\ell}[n]\big\}$ as
\begin{align}
\label{eq:samp_seq}
c_{\ell}[n]=\left\langle y(t),s(t-nT-(\ell-1) T/p) \right\rangle, \ \ell = 1,\dots,p.
\end{align}
These sequences correspond to periodically splitting the samples at the output of the LPF, which is generated at a rate of $p/T$, into $p$ slower sequences at a rate of $1/T$ each using a serial-to-parallel converter; see Fig.~\ref{fig:samp_rec_scheme} for a schematic representation of this splitting.

Next, we define the vector $\Dft{c}{}$ as the $p$-length vector whose $\ell$th element is $\dft{C}{\ell}$, which denotes the \emph{discrete-time Fourier transform} (DTFT) of $c_{\ell}[n]$. In a similar fashion, we define $\Dft{a}{}$ as the $K_{\tau}$-length vector whose $i$th element is given by $\dft{A}{i}$, the DTFT of $a_i[n]$. It can be shown following the developments carried out in \cite{LRTDE} that these two vectors are related to each other by
\begin{align}
\label{eq:c_dtft}
\Dft{c}{}=\Dft{W}{}\mbf{N}\left(\bs{\tau}\right)\mbf{D}\left(e^{j\omega T},\bs{\tau}\right)\Dft{a}{}.
\end{align}
Here, $\Dft{W}{}$ is a $p \times p$ matrix with $(\ell,m)$th element
\begin{align}
\label{eq:Wlm}
\Dft{W}{\ell m}=
e^{j\omega(\ell-1)T/p}
\frac{1}{T} S^*\left(\omega+\frac{2\pi}{T}m'\right)
G\left(\omega+\frac{2\pi}{T}m'\right)
e^{j\frac{2\pi}{p}(\ell-1) m'},
\end{align}
where $m' \defn m-p/2-1$, $\mbf{N}\left(\bs{\tau}\right)$ is a $p \times K_{\tau}$ Vandermonde matrix with $(m,i)$th element
\begin{align}
\label{eq:N_mk}
\mbf{N}_{mi}\left(\bs{\tau}\right)=e^{-j\frac{2\pi}{T}m'\tau_i},
\end{align}
and $\mbf{D}\left(e^{j\omega T},\bs{\tau}\right)$ is a $K_{\tau} \times K_{\tau}$ diagonal matrix whose $i$th diagonal element is given by $e^{-j\omega\tau_i}$. Assuming for the time being that $\Dft{W}{}$ is a stably-invertible matrix, we define the \emph{modified measurement vector} $\Dft{d}{}=\mbf{W}^{-1}\Dft{}{}\Dft{c}{}$. Denoting
\begin{align}
\label{eq:b_dtft}
\Dft{b}{}=\mbf{D}\left(e^{j\omega T},\bs{\tau}\right) \Dft{a}{},
\end{align}
we see from \eqref{eq:Wlm} that
\begin{align}
\label{eq:d_dtft}
\Dft{d}{}=\mbf{N}\left(\bs{\tau}\right)\Dft{b}{}.
\end{align}
Since $\mbf{N}\left(\bs{\tau}\right)$ is not a function of $\omega$, \eqref{eq:d_dtft} can be expressed in the discrete-time domain using the linearity of the DTFT as
\begin{align}
\label{eq:dn}
\mbf{d}[n]=\mbf{N}\left(\bs{\tau}\right)\mbf{b}[n], \quad n \in \mathbb{Z}.
\end{align}
Here, the elements of the vectors $\mbf{d}[n]$ and $\mbf{b}[n]$ are discrete-time sequences that are given by the inverse DTFT of the elements of $\Dft{d}{}$ and $\Dft{b}{}$, respectively.

The key insight to be drawn here is that \eqref{eq:d_dtft}, and its time-domain equivalent \eqref{eq:dn}, can be viewed as an infinite ensemble of modified measurement vectors in which each element corresponds to a distinct matrix $\mbf{N}\left(\bs{\tau}\right)$ that, in turn, depends on the set of (distinct) delays $\bs{\tau}$. Linear measurement models of the form \eqref{eq:dn}---in which the measurement matrix is completely determined by a set of (unknown) parameters---have been studied extensively in a number of research areas such as system identification \cite{moulines1995} and direction-of-arrival and spectrum estimation \cite{krim1996tda,stoica1997}. One specific class of methods that has proven to be quite useful in these areas in efficiently recovering the parameters that describe the measurement matrix are the so-called \emph{subspace methods} \cite{krim1996tda}. Consequently, our approach in the recovery stage will be to first use subspace methods in order to recover the set $\bs{\tau}$ from $\mbf{d}[n]$.  Afterwards, since $\mbf{N}^{\dag}\left(\bs{\tau}\right) \mbf{N}\left(\bs{\tau}\right) = \mbf{I}$ because of the assumption that $p \geq 2K_\tau$, we will recover $\Dft{a}{}$ from $\mbf{d}[n]$ using linear filtering operations as follows [cf.~\eqref{eq:b_dtft}, \eqref{eq:d_dtft}]
\begin{align}
\label{eq:a_recover}
\Dft{a}{}=\mbf{D}^{-1}\left(e^{j\omega T},\bs{\tau}\right)\mbf{N}^{\dag}\left(\bs{\tau}\right)\Dft{d}{}.
\end{align}
Finally, the Doppler-shifts and attenuation factors associated with $\cH$ are determined from the vector $\Dft{a}{}$ by an additional use of the subspace methods.

Before proceeding to the recovery stage, we justify the assumption that $\Dft{W}{}$ can be stably inverted. To this end, observe from \eqref{eq:Wlm} that $\Dft{W}{}$ can be decomposed as
\begin{align}
\label{eq:W_decomp}
\Dft{W}{}=\Dft{\Phi}{}\mbf{F}^{H}\Dft{\Psi}{},
\end{align}
where $\Dft{\Phi}{}$ is a $p \times p$ diagonal matrix with $\ell$th diagonal element
\begin{align}
\Dft{\Phi}{\ell \ell}=\sqrt{p}(-1)^{\ell-1}e^{j\omega(\ell-1)T/p},
\end{align}
$\mbf{F}$ is a $p$-point discrete Fourier transform (DFT) matrix with $(\ell,m)$th element equal to
\begin{align}
\mbf{F}_{\ell m}=\frac{1}{\sqrt{p}}e^{-j\frac{2\pi}{p}(\ell-1)(m-1)},
\end{align}
and $\Dft{\Psi}{}$ is a $p \times p$ diagonal matrix whose $m$th diagonal element is given by
\begin{align}
\Dft{\Psi}{mm}=
\frac{1}{T} S^*\left(\omega+\frac{2\pi}{T}(m-p/2-1)\right)
G\left(\omega+\frac{2\pi}{T}(m-p/2-1)\right).
\end{align}
It can now be easily seen from the decomposition in \eqref{eq:W_decomp} that, in order for $\Dft{W}{}$ to be stably invertible, each of the above three matrices has to be stably invertible. By construction, both $\Dft{\Phi}{}$ and $\mbf{F}^H$ are stably invertible. The invertibility requirement on the diagonal matrix $\Dft{\Psi}{}$ leads to the following conditions on the pulse $g(t)$ and the kernel $s^{*}(-t)$ of the LPF.
\begin{condition}
\label{cond:pulse_cond}
In order for $\Dft{\Psi}{}$ to be stably invertible, the continuous-time Fourier transform of $g\left(t\right)$ has to satisfy
\begin{align}
a\leq\left|G\left(\omega\right)\right|\leq b \quad \textrm{ a.e. }\omega\in\mathcal{F}
\end{align}
for some positive constants $a > 0$ and $b < \infty$.
\end{condition}
\begin{condition}
\label{cond:filter_cond}
In order for $\Dft{\Psi}{}$ to be stably invertible, the continuous-time Fourier transform of the LPF $s^*\left(-t\right)$ has to satisfy
\begin{align}
c \leq\left|S\left(\omega\right)\right|\leq d \quad \textrm{ a.e. }\omega\in\mathcal{F}
\end{align}
for some positive constants $c>0$ and $d<\infty$.
\end{condition}

Condition~\ref{cond:pulse_cond} requires that the bandwidth $\mathcal{W}$ of the prototype pulse $g(t)$ has to satisfy
\begin{align}
\label{eq:min_bw1}
\mathcal{W} \geq \frac{2\pi}{T}p.
\end{align}
In Section~\ref{sec:suff_cond}, we will derive additional conditions on the parameter $p$ and combine them with \eqref{eq:min_bw1} to obtain equivalent requirements on the time--bandwidth product of the input signal $x(t)$ that will ensure invertibility of the matrix $\Dft{W}{}$.

We conclude discussion of the sampling stage by pointing out that the decomposition in \eqref{eq:W_decomp} also provides an efficient way to implement the digital-correction filter bank $\mbf{W}^{-1}\Dft{}{}$. This is because \eqref{eq:W_decomp} implies that
\begin{align}
\label{eqn:w_inv}
\mbf{W}^{-1}\Dft{}{}=\mbf{\Psi}^{-1}\Dft{}{} \mbf{F} \mbf{\Phi}^{-1}\Dft{}{}.
\end{align}
Therefore the implementation of $\mbf{W}^{-1}\Dft{}{}$ can be done in three stages, where each stage corresponds to one of the three matrices in \eqref{eqn:w_inv}. Specifically, define the set of digital filters $\{\phi_{\ell}[n]\}$ and $\{\psi_{\ell}[n]\}$ as
\begin{align}
\phi_{\ell}[n]=\text{IDTFT} \left\{\mbf{\Phi}^{-1}_{\ell \ell}\Dft{}{}\right\}[n], \quad 1\leq \ell \leq p
\end{align}
and
\begin{align}
\psi_{\ell}[n]=\text{IDTFT} \left\{\mbf{\Psi}^{-1}_{\ell \ell}\Dft{}{}\right\}[n], \quad 1\leq \ell \leq p,
\end{align}
where IDTFT denotes the inverse DTFT operation. The first correction stage involves filtering the sequences $\{c_{\ell}[n]\}$ using the set of filters $\{\phi_{\ell}[n]\}$. Next, multiplication with the DFT matrix $\mbf{F}$ can be efficiently implemented by applying the fast Fourier transform (FFT) to the outputs of the filters in the first stage. Finally, the third correction stage involves filtering the resulting sequences using the set of filters $\{\psi_{\ell}[n]\}$ to get the desired sequences $\{d_{\ell}[n]\}$. This last correction stage can be interpreted as an equalization step that compensates for the non-flatness of the frequency responses of the prototype pulse and the kernel of the LPF. A detailed schematic representation of the sampling stage, which is based on the preceding interpretation of $\mbf{W}^{-1}\Dft{}{}$, is provided in Fig.~\ref{fig:samp_rec_scheme}.

\subsection{The Recovery Stage}\label{ssec:recovery}
We conclude this section by describing in detail the recovery stage, which---as noted earlier---consists of two steps. In the first step, we rely on subspace methods to recover the delays $\bs{\tau}$ from $\mbf{d}[n]$ [cf.~\eqref{eq:dn}]. In the second step, we make use of the recovered delays to obtain the Doppler-shifts and attenuation factors associated with each of the delays.

\begin{table}
\caption{Delay Recovery Algorithm}
\label{tab:del_rec}
\hrule
\begin{enumerate}
\item[(i)]
    Construct the matrix
    \begin{align*}
    \overline{\mbf{R}}_{dd}=\frac{1}{M}\sum_{m=1}^{M}
    \sum_{n\in\mathbb{Z}} \mbf{d}_{m}[n] \mbf{d}_{m}^{H}[n],
    \end{align*}
    where $\mbf{d}_{m}$ is the $M=p/2$ length subvector which is given by
    \begin{align*}
    \mbf{d}_{m}\left[n\right]=\left[
    \begin{array}{cccc}
    d_{m}\left[n\right] & d_{m+1}\left[n\right] & \ldots &
    d_{m+M}\left[n\right]
    \end{array}\right]^{T}.
    \end{align*}
\item[(ii)]
    Recover $K_{\tau}$ as the rank of $\overline{\mbf{R}}_{dd}$.
\item[(iii)]
    Perform a \emph{singular value decomposition} (SVD) of $\overline{\mbf{R}}_{dd}$ and construct the matrix $\mbf{E}_{s}$ consisting of the $K_{\tau}$ singular vectors corresponding to the $K_\tau$ nonzero singular values of $\overline{\mbf{R}}_{dd}$ as its columns.
\item[(iv)]
    Compute the matrix $\mbf{\Phi=\mbf{E}_{s\downarrow}^{\dagger}}\mbf{\mbf{E}_{s\uparrow}}$, where $\mathbf{E}_{s\uparrow}$ and $\mathbf{E}_{s\downarrow}$ denote the submatrices extracted from $\mathbf{E}_{s}$ by removing its first row and its last row, respectively.
\item[(v)]
    Compute the eigenvalues of $\mbf{\Phi}$, $\lambda_{i},i=1,2,\ldots,K_{\tau}$.
\item[(vi)]
    Recover the unknown delays as follows: $\tau_{i}=-\frac{T}{2\pi}\textrm{arg}\left(\lambda_{i}\right)$.
\end{enumerate}
\hrule
\end{table}

\subsubsection{Recovery of the Delays}
\label{sub:delays_rec}
In order to recover $\bs{\tau}$ from $\mbf{d}[n]$, we rely on the approach advocated in \cite{LRTDE} and make use of the well-known ESPRIT algorithm \cite{32276} together with an additional smoothing stage \cite{smooth}. The exact algorithm is given in Table~\ref{tab:del_rec}; we refer the reader to \cite{LRTDE,32276} for details.

\subsubsection{Recovery of the Doppler-Shifts and Attenuation Factors}
\label{sub:dopplers_rec}
Once the unknown delays are found, we can recover the vectors $\mbf{a}[n]$
through the frequency relation \eqref{eq:a_recover}. Next, define for each delay $\tau_i$, the set of corresponding Doppler-shifts
\begin{align}
\bs{\nu}_i \defn \{ \nu_{ij}, \ j=1,\dots,K_{\nu,i}\}
\end{align}
and recall that the $i$th element of $\mbf{a}[n]$ is given by \eqref{eq:sequences}. We can therefore write the $N$-length sequence $\{a_i[n]\}$ for each index $i$ in the following matrix--vector form
\begin{align}
\label{eq:a_k}
\mbf{a}_i=\mbf{X}\mbf{R}(\bs{\nu}_i)\bs{\alpha}_i,
\end{align}
where $\mbf{a}_i$ is a length-$N$ vector whose $n$th element is $a_i[n]$, $\mbf{X}$ is an $N \times N$ diagonal matrix whose $n$th diagonal element is given by $x_n$, $\mbf{R}(\bs{\nu}_i)$ is an $N \times K_{\nu,i}$ Vandermonde matrix with $(n,j)$th element $e^{j2\pi \nu_{ij} nT}$, and $\bs{\alpha}_i$ is length-$K_{\nu,i}$ vector with $j$th element $\alpha_{ij}$. The matrix $\mbf{X}$ in \eqref{eq:a_k} can be inverted under the assumption that the sequence $\{x_n\}$ satisfies $|x_n| > 0$ for every $n=0,\dots,N-1$, resulting in
\begin{align}
\label{eq:a_k_tilde}
\tilde{\mbf{a}}_i=\mbf{R}(\bs{\nu}_i)\bs{\alpha}_i,
\end{align}
where $\tilde{\mbf{a}}_i \defn \mbf{X}^{-1}\bs{a}_i$. From a simple inspection, we can express the elements of $\tilde{\mbf{a}}_i$ as
\begin{align}
\label{eq:sequences2}
\tilde{a}_i[n] = \sum_{j=1}^{K_{\nu,i}}\alpha_{ij} e^{j2\pi \nu_{ij} nT},\quad 0\leq n \leq N-1.
\end{align}
It is now easy to see from this representation that recovery of the Doppler-shifts from the sequences $\{\tilde{a}_i[n]\}$ is equivalent to the problem of recovering distinct frequencies from a (weighted) sum of complex exponentials. In the context of our problem, for each fixed index $i$, the frequency of the $j$th exponential is $\omega_{i j}=2\pi \nu_{ij} nT$ and its amplitude is $\alpha_{ij}$.

Fortunately, the problem of recovering frequencies from a sum of complex exponentials has been studied extensively in the literature and various strategies exist for solving this problem (see \cite{stoica1997} for a review). One of these techniques that has gained interest recently, especially in the literature on finite rate of innovation \cite{4156380,1003065,ronen_kfir,RonenSingleChannel}, is the \emph{annihilating-filter} method. The annihilating-filter approach, in contrast to some of the other techniques, allows the recovery of the frequencies associated with the $i$th index even at the critical value of $N = 2 K_{\nu,i}$. On the other hand, subspace methods such as ESPRIT \cite{1164935}, matrix-pencil algorithm \cite{hua1990matrix}, and the Tufts and Kumaresan approach \cite{KT} are generally more robust to noise but also require more samples than $2 K_{\nu,i}$. Once the Doppler-shifts for each index $i$ have been recovered then, since $\mbf{R}^{\dag}(\bs{\nu}_i)\mbf{R}(\bs{\nu}_i) = \mbf{I}$ because of the requirement that $N \geq 2K_{\nu,i}$, the attenuation factors $\{\alpha_{ij}\}$ are determined as
\begin{align}
\label{eq:reflection_rec}
\bs{\alpha}_i=\mbf{R}^{\dag}(\bs{\nu}_i)\tilde{\mbf{a}}_i, \quad i=1,\dots,K_\tau.
\end{align}

%%%%%%%%%%%%%%%%%%%%%%%%%%%%%%%%%%%%%%%%%%%%%%%%%%%%%%%%%%%%%
\section{Sufficient Conditions for Identification}\label{sec:suff_cond}
Our focus in Section~\ref{sec:id_sch} was on developing a recovery algorithm for the identification of ULSs. We now turn to specify conditions that guarantee that the proposed procedure recovers the set of triplets $\big\{(\tau_i, \nu_{ij}, \alpha_{ij})\big\}$ that describe the parametric ULS $\cH$. We present these requirements in terms of equivalent conditions on the time--bandwidth product $\cT \cW$ of the input signal $x(t)$. This is a natural way to describe the performance of system identification schemes since $\cT \cW$ roughly defines the number of temporal degrees of freedom available for estimating $\cH$ \cite{slepian:pieee76}.

To begin with, recall that the recovery stage involves first determining the unknown delays $\bs{\tau}$ from the set of equations given by \eqref{eq:dn} (cf.~Section~\ref{ssec:recovery}). Therefore, to ensure that our algorithm successfully returns the parameters of $\cH$, we first need to provide conditions that guarantee a unique solution to \eqref{eq:dn}. To facilitate the forthcoming analysis, we let $\mbf{d}\left[\Lambda\right]\defn\left\{ \mbf{d}\left[n\right],n\in\mathbb{Z}\right\}$ and $\mbf{b}\left[\Lambda\right]\defn\left\{ \mbf{b}\left[n\right],n\in\mathbb{Z}\right\}$ denote the set of all vectors $\mbf{d}\left[n\right]$ and $\mbf{b}\left[n\right]$, respectively. Using this notation, we can rewrite \eqref{eq:dn} as
\begin{align}
\label{eq:d_Lambda}
\mbf{d}\left[\Lambda\right]=\mbf{N}\left(\bs{\tau}\right)\mbf{b}\left[\Lambda\right].
\end{align}
We now leverage the analysis carried out in \cite{LRTDE} to provide sufficient conditions for a unique solution to \eqref{eq:d_Lambda}; see \cite{LRTDE} for a formal proof.
\begin{proposition}
\label{pro:unique}
If $\left(\bar{\bs{\tau}},\bar{\mathbf{b}}\left[\Lambda\right] \neq \mathbf{0} \right)$ solves \eqref{eq:d_Lambda} and if
\begin{align}
p>2K_{\tau}-\textrm{dim}\left(\textrm{span}\left(\bar{\mathbf{b}}\left[\Lambda\right]\right)\right)\label{eq:p>2K-dim}
\end{align}
then $\left(\bar{\bs{\tau}},\bar{\mathbf{b}}\left[\Lambda\right]\right)$ is the unique solution of \eqref{eq:d_Lambda}. Here, $\textrm{span}\left(\bar{\mbf{b}}\left[\Lambda\right]\right)$ is used to denote the subspace of minimal dimensions that contains $\bar{\mbf{b}}\left[\Lambda\right]$.
\end{proposition}

Proposition~\ref{pro:unique} suggests that a unique solution to \eqref{eq:d_Lambda}---and, by extension, unique recovery of the set of delays $\bs{\tau}$---is guaranteed through a proper selection of the parameter $p$. In particular, since $\textrm{dim}\left(\textrm{span}\left(\bar{\mbf{b}}\left[\Lambda\right]\right)\right)$ is a positive number in general, we have from Proposition~\ref{pro:unique} that $p \geq 2K_{\tau}$ is a sufficient condition for unique recovery of $\bs{\tau}$ and ${\mbf{b}}\left[\Lambda\right]$. From Condition~\ref{cond:pulse_cond} in Section~\ref{sec:id_sch}, we have seen that the parameter $p$ effectively determines the minimum bandwidth $\cW$ of the prototype pulse [cf.~\eqref{eq:min_bw1}]. Combining the condition $p \geq 2K_{\tau}$ and the one obtained earlier in \eqref{eq:min_bw1} leads to the following sufficient condition on the bandwidth of the input signal for unique recovery of $\bs{\tau}$ and ${\mbf{b}}\left[\Lambda\right]$:
\begin{align}
\label{eq:min_bw2}
\mathcal{W} \geq \frac{4\pi K_{\tau}}{T}.
\end{align}

The second step in the recovery stage involves recovering the Doppler-shifts and attenuation factors (cf.~Section~\ref{ssec:recovery}). We now investigate the conditions required for unique recovery of the Doppler-shifts. Recall that the vectors $\mbf{b}[n]$ and $\mbf{a}[n]$ are related to each other by the invertible frequency relation \eqref{eq:b_dtft}. Therefore, since the diagonal matrix $\mbf{D}\left(e^{j\omega T},\bs{\tau}\right)$ is invertible and completely specified by $\bs{\tau}$, the condition given in \eqref{eq:min_bw2} also guarantees unique recovery of the vectors $\mbf{a}[n]$ from $\mbf{b}[n]$. Further, it can be easily verified that the matrix $\mbf{R}(\bs{\nu}_i)$ in \eqref{eq:a_k_tilde} has the same parametric structure as that required by Proposition~\ref{pro:unique}. We can therefore once again appeal to the results of Proposition~\ref{pro:unique} in providing conditions for unique recovery of the Doppler-shifts $\{\bs{\nu}_i\}$ from the vectors $\{\tilde{\mbf{a}}_i\}$ [cf.~\eqref{eq:a_k_tilde}]. To that end, we interchange $p$ with $N$ and $K_{\tau}$ with $K_{\nu,i}$ in Proposition~\ref{pro:unique} and use the fact that $\textrm{dim}\left(\textrm{span}\left(\bs{a}_i\right)\right)=1$ (since it is a nonzero vector). Therefore, by making use of Proposition~\ref{pro:unique}, we conclude that a sufficient condition for unique recovery of $\bs{\nu}_i$ from \eqref{eq:a_k_tilde} is
\begin{align}
\label{eq:N_cond1}
N \geq 2K_{\nu,i}.
\end{align}

Condition \eqref{eq:N_cond1} is intuitive in the sense that there are $2K_{\nu,i}$ unknowns in \eqref{eq:a_k_tilde} ($K_{\nu,i}$ unknown Doppler-shifts and $K_{\nu,i}$ unknown attenuation factors) and therefore at least $2K_{\nu,i}$ equations are required to solve for these unknown parameters. Finally, since we need to ensure unique recovery of the Doppler-shifts and attenuation factors for each distinct delay $\tau_i$, we have the condition
\begin{align}
N \geq 2 \max_i K_{\nu,i}
\end{align}
which trivially ensures that \eqref{eq:N_cond1} holds for every $i=1,\ldots,K_{\tau}$. We summarize these results in the following theorem.
\begin{theorem}[Sufficient Conditions for System Identification]
\label{thm:summ_results}
Suppose that $\cH$ is a parametric ULS that is completely described by a total of $K = \sum_{i=1}^{K_\tau} K_{\nu,i}$ triplets $(\tau_i, \nu_{ij}, \alpha_{ij})$. Then, irrespective of the distribution of $\{(\tau_i, \nu_{ij})\}$ within the delay--Doppler space, the recovery procedure specified in Section~\ref{sec:id_sch} with samples taken at $\{t = 2n\pi/\cW\}$ uniquely identifies $\cH$ from a single observation $\cH(x(t))$ as long as the system satisfies \textbf{[A1]}--\textbf{[A3]}, the probing sequence $\{x_n\}$ remains bounded away from zero in the sense that $|x_n| > 0$ for every $n=0,\dots,N-1$, and the time--bandwidth product of the (known) input signal $x(t)$ satisfies the condition
\begin{align}
 \mathcal{TW} \geq 8 \pi K_\tau K_{\nu,max}
\end{align}
where $K_{\nu,max} \defn \max_i K_{\nu,i}$ is the maximum number of Doppler-shifts associated with any one of the distinct delays.
\end{theorem}
\begin{proof}
Recall from the previous discussion that the delays, Doppler-shifts, and attenuation factors associated with $\cH$ can be uniquely recovered as long as $N \geq 2K_{\nu,max}$, $\mathcal{W} \geq \frac{4\pi K_{\tau}}{T}$, and $p \geq 2 K_\tau$. Now take $N = \frac{\mathcal{TW}}{4 \pi K_\tau}$ and note that under the assumption $\mathcal{TW} \geq 8 \pi K_\tau K_{\nu,max}$, we trivially have $N \geq 2K_{\nu,max}$. Further, since $\cT = NT$ and since sampling rate of $2\pi/\cW$ implies $p = 2\pi/ (\cW T)$, we also have that $\mathcal{W} = \frac{4\pi K_{\tau}}{T} \ \Rightarrow \ p = 2K_\tau$, completing the proof.
\end{proof}

Theorem~\ref{thm:summ_results} implicitly assumes that $K_\tau$ (or an upper bound on $K_\tau$) and $K_{\nu,max}$ (or an upper bound on $K_{\nu,max}$) are known at the transmitter side. We explore this point in further detail in Section~\ref{sec:disc} and numerically study the effects of ``model-order mismatch'' on the robustness of the proposed recovery procedure. It is also instructive (especially for comparison purposes with related work such as \cite{bajwa:allerton08,herman:tsp09}) to present a weaker version of Theorem~\ref{thm:summ_results} that only requires knowledge of the total number of delay--Doppler pairs $K$.
\begin{corollary}[Weaker Sufficient Conditions for System Identification]\label{thm:suff_cond_rpt}
Suppose that the assumptions of Theorem~\ref{thm:summ_results} hold. Then the recovery procedure specified in Section~\ref{sec:id_sch} with samples taken at $\{t = 2n\pi/\cW\}$ uniquely identifies $\cH$ from a single observation $\cH(x(t))$ as long as the time--bandwidth product of the known input signal $x(t)$ satisfies the condition $\cT\cW \geq 2 \pi (K+1)^2$.
\end{corollary}
\begin{proof}
This corollary is a simple consequence of Theorem~\ref{thm:summ_results} and the fact that $K_\tau K_{\nu,max} \leq \frac{(K+1)^2}{4}$. To prove the latter fact, note that for any fixed $K$ and $K_\tau$, we always have $K_{\nu,max} \leq K - (K_\tau - 1)$. Indeed, if $K_{\nu,max}$ were greater than $K - (K_\tau - 1)$ then either $\sum_{i=1}^{K_\tau} K_{\nu,i} > K$ or there exists an $i$ such that $K_{\nu,i} = 0$, both of which are contradictions. Consequently, for any fixed $K$, we have that
\begin{align}
	K_\tau K_{\nu,max} \leq - K_\tau^2 + (K+1)K_\tau
\end{align}
and since the maximum of $- K_\tau^2 + (K+1)K_\tau$ occurs at $K_\tau = \frac{K+1}{2}$, we get $K_\tau K_{\nu,max} \leq \frac{(K+1)^2}{4}$.
\end{proof}

%%%%%%%%%%%%%%%%%%%%%%%%%%%%%%%%%%%%%%%%%%%%%%%%%%%%%%%%%%%%%
\section{Discussion}\label{sec:disc}
In Sections~\ref{sec:id_sch} and \ref{sec:suff_cond}, we proposed and analyzed a polynomial-time recovery procedure that ensures identification of parametric ULSs under certain conditions. In particular, one of the key contributions of the preceding analysis is that it parlays a key sub-Nyquist sampling result of \cite{LRTDE} into conditions on the time--bandwidth product, $\cT \cW$, of the input signal $x(t)$ that guarantee identification of arbitrary linear systems as long as they are sufficiently underspread. Specifically, in the parlance of system identification, Corollary~\ref{thm:suff_cond_rpt} states that the recovery procedure of Section~\ref{sec:id_sch} achieves \emph{infinitesimally-fine resolution} in the delay--Doppler space as long as the temporal degrees of freedom available to excite a ULS are on the order of $\Omega(K^2)$. In addition, we carry out extensive numerical experiments in Section~\ref{sec:num_res}, which confirm that---as long as the condition $\cT\cW \geq 2 \pi (K+1)^2$ is satisfied---the ability of the proposed procedure to distinguish between (resolve) closely spaced delay--Doppler pairs is primarily a function of the signal-to-noise ratio (SNR) and its performance degrades gracefully in the presence of noise. In order to best put the significance of our results into perspective, it is instructive to compare them with corresponding results in recent literature. We then discuss an application of these results to super-resolution target detection using radar in Section~\ref{sec:radar}.

There exists a large body of existing work---especially in the communications and radar literature---treating identification of parametric ULSs; see, e.g., \cite{kay:icassp97, jakobsson:tsp98, skolnik:01, wong:globecom06, bajwa:allerton08, herman:tsp09,tan:tsp09}. One of the approaches that is commonly taken in many of these works, such as in \cite{kay:icassp97,wong:globecom06,bajwa:allerton08,herman:tsp09,tan:tsp09}, is to quantize the delay--Doppler space $(\tau,\nu)$ by assuming that both $\tau_i$ and $\nu_{ij}$ lie on a grid. The following theorem is representative of some of the known results in this case.\footnote{It is worth mentioning here that a somewhat similar result was also obtained independently in \cite{pfander:jfaa09} in an abstract setting.}
\begin{theorem}[\!\!\cite{bajwa:allerton08,herman:tsp09}]\label{thm:disc_cond}
Suppose that $\cH$ is a parametric ULS that is completely described by a total of $K = \sum_{i=1}^{K_\tau} K_{\nu,i}$ triplets $(\tau_i, \nu_{ij}, \alpha_{ij})$. Further, let the delays and the Doppler-shifts of the system be such that $\tau_i = r_i \cW^{-1}$ and $\nu_{ij} = \ell_{ij} \cT^{-1}$ for $r_i \in \mathbb{Z}_+$ and $\ell_{ij} \in \mathbb{Z}$. Then $\cH$ can be identified in polynomial-time from a single observation $\cH(x(t))$ as long as the system satisfies \textbf{[A1]}--\textbf{[A3]} and the time--bandwidth product of the input signal $x(t)$ satisfies $\cT\cW = \Omega(K^2/\log{\cT\cW})$.
\end{theorem}

There are two conclusions that can be immediately drawn from Theorem~\ref{thm:disc_cond}. First, both \cite{bajwa:allerton08,herman:tsp09} require about the same scaling of the temporal degrees of freedom as that required by Corollary~\ref{thm:suff_cond_rpt}: $\cT\cW \approx \Omega(K^2)$. Second, the resolution of the recovery procedures proposed in \cite{bajwa:allerton08,herman:tsp09} is limited to $\cW^{-1}$ in the delay space and $\cT^{-1}$ in the Doppler space because of the assumption that $\tau_i = r_i \cW^{-1}$ and $\nu_{ij} = \ell_{ij} \cT^{-1}$.\footnote{Note that there is also a Bayesian variant of Theorem~\ref{thm:disc_cond} in \cite{herman:tsp09} that requires $\cT\cW \approx \Omega(K)$ under the assumption that $\cH$ has a uniform statistical prior over the quantized delay--Doppler space. A somewhat similar Bayesian variant of Corollary~\ref{thm:suff_cond_rpt} can also be obtained by trivially extending the results of this paper to the case when $\cH$ is assumed to have a uniform statistical prior over the \emph{non-quantized} delay--Doppler space.} Similarly, in another related recent paper \cite{tan:tsp09}, two recovery procedures are proposed that have been numerically shown to uniquely identify $\cH$ as long as $\cT\cW \gg 1$ and each $(\tau_i, \nu_{ij})$ corresponds to one of the points in the quantized delay--Doppler space with resolution proportional to $\cW^{-1}$ and $\cT^{-1}$ in the delay space and the Doppler space, respectively. Note that the assumption of a quantized delay--Doppler space can have unintended consequences in certain applications and we carry out a detailed discussion of this issue in the next section in the context of radar target detection.

Finally, the work in \cite{jakobsson:tsp98} leverages some of the results in DOA estimation to propose a scheme for the identification of linear systems of the form \eqref{eqn:LTV_sys} without requiring that $\tau_i = r_i \cW^{-1}$ and $\nu_{ij} = \ell_{ij} \cT^{-1}$. Nevertheless, our results differ from those in \cite{jakobsson:tsp98} in three important respects. First, we explicitly state the relationship between the time--bandwidth product $\cT \cW$ of the input signal $x(t)$ and the number of delay--Doppler pairs $K = \sum_{i=1}^{K_\tau} K_{\nu,i}$ that guarantees recovery of the system response by studying the sampling and recovery stages of our proposed recovery procedure. On the other hand, the method proposed in \cite{jakobsson:tsp98} assumes the sampling stage to be given and, as such, fails to make explicit the connection between the time--bandwidth product of $x(t)$ and the number of delay-Doppler pairs. Second, the algorithms proposed in \cite{jakobsson:tsp98} have exponential complexity, since they require exhaustive searches in a $K$-dimensional space, which can be computationally prohibitive for large-enough values of $K$. Last, but not the least, recovery methods proposed in \cite{jakobsson:tsp98} are guaranteed to work as long as there are no more than two delay--Doppler pairs having the same delay, $\max_i K_{\nu,i} \leq 2$, and the system output is observed by an $M$-element antenna array with $M \gtrapprox K$. In contrast, our recovery algorithm does not impose any restrictions on the distribution of $\{(\tau_i, \nu_{ij})\}$ within the delay-Doppler space and is guaranteed to work with a single observation of the system output.

%%%%%%%%%%%%%%%%%%%%%%%%%%%%%%%%%%%%%%%%%%%%%%%%%%%%%%%%%%%%%
\section{Application: Super-Resolution Radar}\label{sec:radar}
\begin{figure}
\begin{center}
\includegraphics[scale=0.355, trim=35 15 0 0]{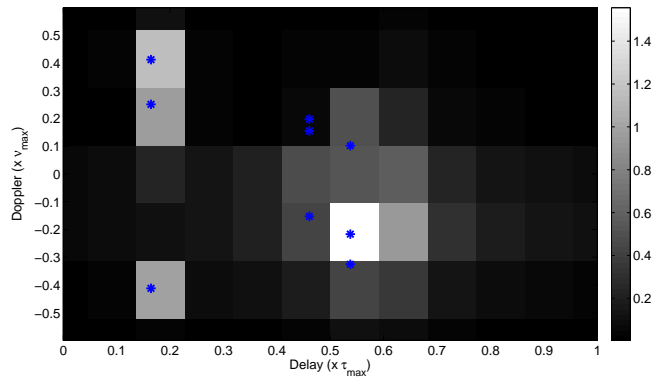}
\caption{\label{fig:discrete}Quantized representation of nine targets (represented by $*$) in the delay--Doppler space with $\tau_{max} = 10~\mu\text{s}$ and $\nu_{max} = 10 \text{ kHz}$. The quantized delay--Doppler approximation of the targets corresponds to $\cW = 1.2$ MHz and $\cT = 0.48$ ms.}
\end{center}
\end{figure}
We have established in Section~\ref{sec:suff_cond} that the polynomial-time recovery procedure of Section~\ref{sec:id_sch} achieves infinitesimally-fine resolution in the delay--Doppler space under mild assumptions on the temporal degrees of freedom of the input signal. This makes the proposed algorithm extremely useful for application areas in which the system performance depends critically on the ability to resolve closely spaced delay--Doppler pairs. In particular, our method can be used for super-resolution target detection using radar. This is because the noiseless, clutter-free received signal in the case of monostatic radars is exactly of the form \eqref{eqn:LTV_sys1} with each triplet $(\tau_k, \nu_k, \alpha_k)$ corresponding to an echo of the radar waveform $x(t)$ from a distinct target \cite{skolnik:01}.\footnote{In the radar literature, the term ``monostatic'' refers to the common scenario of the radar transmitter and the radar receiver being collocated.} The fact that our recovery procedure allows to identify arbitrary parametric ULSs, therefore, enables us to distinguish between multiple targets even if their radial positions are quite close to each other and/or their radial velocities are similar---the so-called super-resolution detection of targets.

On the other hand, note that apart from the fact that none of the methods based on the assumption of a quantized delay--Doppler space can ever carry out super-resolution target detection, a major drawback of the radar target detection approach in works such as \cite{herman:tsp09, tan:tsp09} is that targets in the real-world do not in fact correspond to points in the quantized delay--Doppler space, which causes \emph{leakage} of their energies in the quantized space. In order to elaborate further on this point, define $L \defn \lceil W\tau_{max} \rceil$ and $M \defn \lceil T\nu_{max}/2 \rceil$ and note that (canonical) quantization corresponds to transforming the $\mathcal{C} = [0,\tau_{max}]\times[-\nu_{max}/2,\nu_{max}/2]$ continuous delay--Doppler space into a $\mathcal{Q} \defn \{0,\dots,L\}\times\{-M/2,\dots,M/2\}$ two-dimensional quantized grid, which in turn transforms the received signal $\cH(x(t))$ at the radar into \cite[Chapter~4]{bello:tc63,bajwa:thesis}
\begin{align}
\label{eqn:quantized_dd}
	\tilde{\cH}(x(t)) \approx \sum_{\ell=0}^{L} \sum_{m=-M}^{M} \tilde{\alpha}_{\ell m} x(t - \tilde{\tau}_{\ell}) e^{j 2 \pi \tilde{\nu}_{m} t}
\end{align}
where $\tilde{\alpha}_{\ell m} \defn \sum_{i=1}^{K_\tau} \sum_{j=1}^{K_{\nu,i}} \alpha_{ij} e^{j \pi (m-\mathcal{T} \nu_{ij})} \text{sinc}(m-\mathcal{T} \nu_{ij}) \text{sinc}(\ell-T \mathcal{W} \tau_{i})$ and the quantized delay--Doppler pairs $(\tilde{\tau}_\ell, \tilde{\nu}_{m}) \in \mathcal{Q}$. It is now easy to conclude from \eqref{eqn:quantized_dd} that, unless the original targets (delay--Doppler pairs) happen to lie in $\mathcal{Q}$, most of the attenuation factors $\{\tilde{\alpha}_{\ell m}\}$ will be nonzero because of the sinc kernels---the so-called ``leakage effect.'' This has catastrophic implications for target detection using radar since leakage makes it impossible to reliably identify the original set of delays and Doppler-shifts. This limitation of target-detection methods that are based on the assumption of a quantized delay--Doppler space is also depicted in Fig.~\ref{fig:discrete} for the case of nine hypothetical targets. The figure illustrates that each of the nine non-quantized targets not only contributes energy to its own $(\tilde{\tau}_\ell, \tilde{\nu}_{m})$ in $\mathcal{Q}$ but also leaks its energy to the nearby points in the quantized space.

\begin{figure}
\centering%
\subfigure[]{\includegraphics[scale=\FIGSCALE, trim=35 0 0 0]{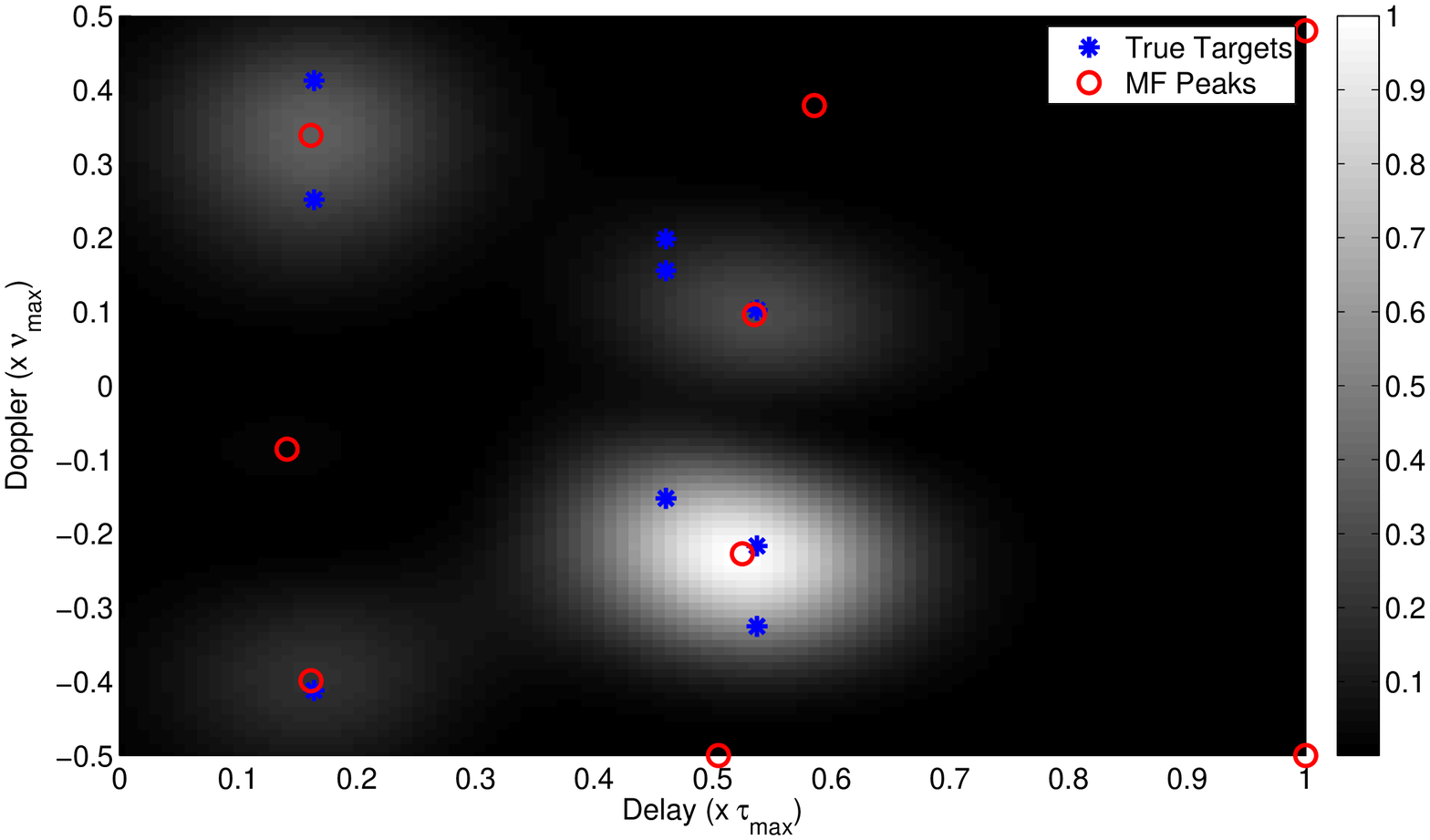}}%
\hfill%
\subfigure[]{\includegraphics[scale=\FIGSCALE, trim=35 0 0 0]{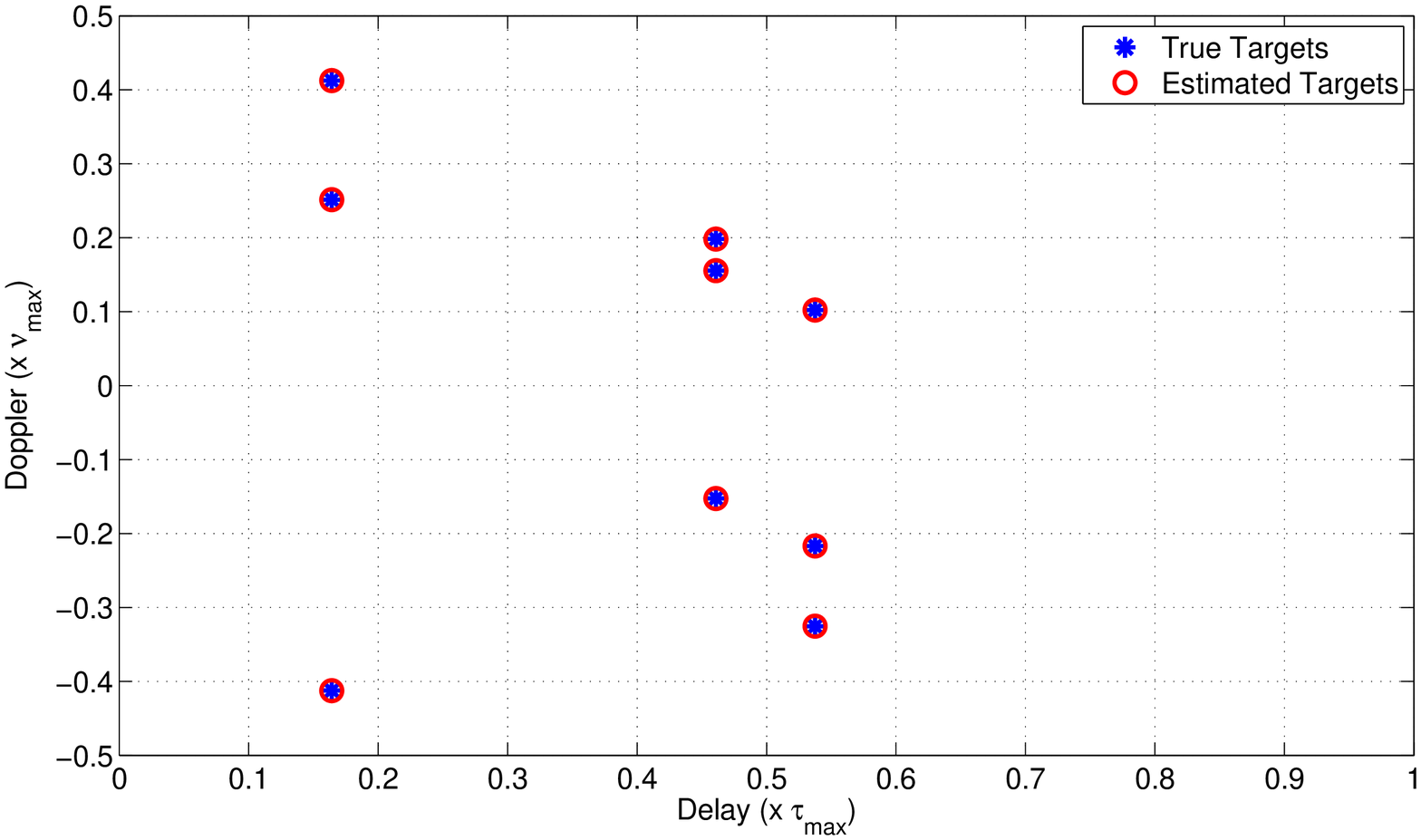}}%
\caption{Comparison between the target-detection performance of matched-filtering and our proposed recovery procedure for the case of nine targets (represented by $*$) in the delay--Doppler space with $\tau_{max} = 10~\mu\text{s}$, $\nu_{max} = 10 \text{ kHz}$, $\cW = 1.2$ MHz, and $\cT = 0.48$ ms. The sequence $\{x_n\}$ corresponds to a random binary $(\pm 1)$ sequence with $N = 48$, the pulse $g(t)$ is designed to have a nearly-flat frequency response in the working spectral band $\m{F}$, and the pulse repetition interval is taken to be $T = 10~\mu\text{s}$. (a) Target detection by matched-filtering the received signal $\cH(x(t))$ with the input signal $x(t)$. (b) Target detection using the proposed recovery procedure with $p = 12$.}%
\label{fig:matched_filter}%
\end{figure}
Owing to the fact that leakage can cause missed detections and false alarms, conventional radar literature in fact tends to focus only on recovery procedures that do no impose any structure on the distribution of $\{(\tau_i, \nu_{ij})\}$ within the delay--Doppler space. The most commonly used approach in the radar signal processing literature corresponds to matched-filtering (MF) the received signal with the input signal $x(t)$ in the delay--Doppler space \cite{skolnik:01}. The MF output $\chi(\tau,\nu)$ takes the form
\begin{align}
\label{eqn:mf_op}
 \chi(\tau,\nu) \defn \int_\mathbb{R} \cH(x(t)) x^*(t - \tau) \exp(-j2\pi \nu t) dt = \sum_{i=1}^{K_\tau} \sum_{j=1}^{K_{\nu,i}} \alpha_{ij} \mathcal{A}(\tau - \tau_i, \nu - \nu_{ij})
\end{align}
where $\mathcal{A}(\tau, \nu) \defn \int_\mathbb{R} x(t) x^*(t-\tau) \exp(-j2\pi \nu t) dt$ is termed the Woodward's \emph{ambiguity function} of $x(t)$. It can be easily deduced from \eqref{eqn:mf_op} that the resolution of the MF-based recovery procedure is tied to the \emph{support} of the ambiguity function in the delay--Doppler space. Ideally, one would like to have $\mathcal{A}(\tau, \nu) = \delta(\tau)\delta(\nu)$ for super-resolution detection of targets but two fundamental properties of ambiguity functions, namely, $|\mathcal{A}(0, 0)|^2 = \int|x(t)|^2dt$ and $\int\!\!\int|\mathcal{A}(\tau, \nu)|^2 d\tau d\nu = \int|x(t)|^2dt$, dictate that no real-world signal $x(t)$ can yield infinitesimally-fine resolution in this case either \cite{skolnik:01}. In fact, the resolution of MF-based recovery techniques also tends to be on the order of $\cW^{-1}$ and $\cT^{-1}$ in the delay space and the Doppler space, respectively, which severely limits their ability to distinguish between two closely-spaced targets in the delay--Doppler space. This inability of MF-based methods to resolve closely-spaced delay--Doppler pairs is depicted in Fig.~\ref{fig:matched_filter}. This figure compares the target-detection performance of MF and the recovery procedure proposed in this paper for the case of nine closely-spaced targets. It is easy to see from Fig.~\ref{fig:matched_filter}(a) that matched-filtering the received signal $\cH(x(t))$ with the input signal $x(t)$ gives rise to peaks that are not centered at the true targets for a majority of the targets. On the other hand, Fig.~\ref{fig:matched_filter}(b) illustrates that our recovery procedure correctly identifies the locations of all nine of the targets in the delay--Doppler space.

%%%%%%%%%%%%%%%%%%%%%%%%%%%%%%%%%%%%%%%%%%%%%%%%%%%%%%%%%%%%%
\section{Numerical Experiments}\label{sec:num_res}
\begin{figure}
\begin{center}
\includegraphics[scale=0.355, trim=35 15 0 0]{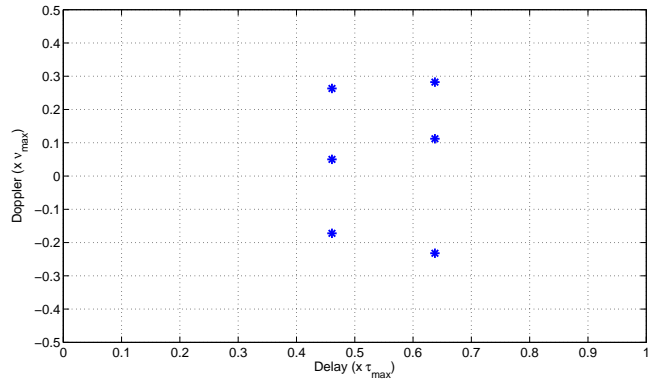}
\caption{\label{fig:dd_plane}Delay--Doppler representation of a parametric ULS $\cH$ corresponding to $K=6$ delay--Doppler pairs with $\tau_{max} = 10~\mu\text{s}$ and $\nu_{max} = 10 \text{ kHz}$.}
\end{center}
\end{figure}
In this section, we explore various issues using numerical experiments that were not treated theoretically earlier in the paper. These include robustness of our method in the presence of noise and the effects of truncated digital filters, use of finite number of samples, choice of probing sequence $\{x_n\}$, and model-order mismatch on the recovery performance. Throughout this section, the numerical experiments correspond to a parametric ULS $\cH$ that is described by a total of $K=6$ delay--Doppler pairs with $K_{\tau}=2$ and $K_{\nu,1} = K_{\nu,2} = 3$. The locations of these pairs in the delay--Doppler space are given by Fig.~\ref{fig:dd_plane}, while the attenuation factors associated with each of the six delay--Doppler pairs are taken to have unit amplitudes and random phases.

In order to identify $\cH$, we design the prototype pulse $g(t)$ to have a constant frequency response over the working spectral band $\m{F} = \left[-\frac{\pi}{T}p,\frac{\pi}{T}p\right]$ with $p = 4$ and $T = 10~\mu\text{s}$, that is, $G(\omega) \approx 1$ when $\omega \in \m{F}$ and $G(\omega) \approx 0$ when $\omega \notin \m{F}$. In other words, the input signal $x(t)$ is chosen to have bandwidth $\mathcal{W}=\frac{8\pi}{T}$. In addition, unless otherwise noted, we use a probing sequence $\{x_n\}$ corresponding to a random binary $(\pm 1)$ sequence with $N = 30$, which leads to a time--bandwidth product of $\mathcal{TW} \approx 240\pi$. Note that the chosen time--bandwidth product here is more than the lower bound of Theorem~\ref{thm:summ_results} by a factor of $5$ so as to increase the robustness to noise. Also, unless otherwise stated, all experiments in the following use an ideal (flat) LPF as the sampling filter (cf.~Fig.~\ref{fig:samp_rec_scheme}). We use the ESPRIT method described in Section~\ref{sec:id_sch} for recovery of the delays and the matrix-pencil method \cite{hua1990matrix} for recovery of the corresponding Doppler-shifts. Given the rich history of these two subspace methods, there exist many standard techniques in the literature (see, e.g., \cite{wax:thesis, roy:thesis}) for providing them with reliable estimates of the model orders in the presence of noise. As such, we assume in the following that both these methods have access to correct values of $K_\tau$ and $K_{\nu,i}$'s. Finally, the performance metrics that we use in this section are the (normalized) mean-squared error (MSE) of the estimated delays and Doppler-shifts (averaged over $100$ noise realizations), defined as
\begin{align}
  e^2_{delay} &= \frac{1}{2}\sum_{i=1}^{2}\big[(\hat{\tau}_{i}-\tau_{i})/\tau_{max}\big]^2,
\intertext{and}
  e^2_{Doppler} &= \frac{1}{6}\sum_{i=1}^{2}\sum_{j=1}^{3}\big[(\hat{\nu}_{ij}-\nu_{ij})/\nu_{max}\big]^2,
\end{align}
where $\hat{\tau}_{i}$ and $\hat{\nu}_{ij}$ denote the estimated delays and Doppler-shifts, respectively.

\begin{figure}
\begin{center}
\includegraphics[scale=0.355, trim=35 15 0 0]{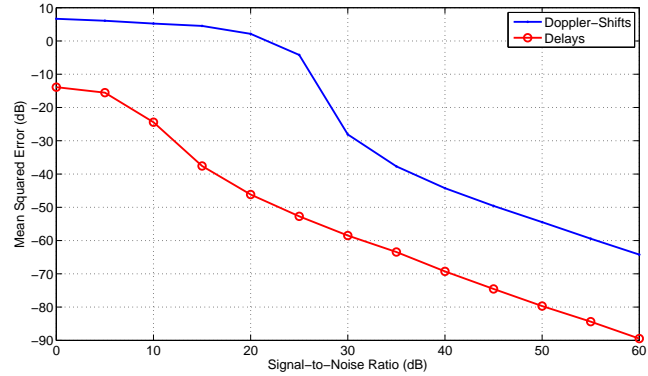}
\caption{\label{fig:VarSNR}Mean-squared error of the estimated delays and Doppler-shifts as a function of the signal-to-noise ratio.}
\end{center}
\end{figure}

%
% hacking ...
\setcounter{subsubsection}{0}%
% End of hacking
%
\subsubsection{Robustness to Noise}
We first examine the robustness of our method when the received signal $\cH(x(t))$ is corrupted by additive noise. The results of this experiment are shown in Fig.~\ref{fig:VarSNR}, which plots the MSE of the estimated delays and Doppler-shifts as a function of the SNR. It can be seen from the figure that the ability of the proposed procedure to resolve delay--Doppler pairs is primarily a function of the SNR and its performance degrades gracefully in the presence of noise.

\subsubsection{Effects of Truncated Digital-Correction Filter Banks}
\begin{figure}
\centering%
\subfigure[]{\includegraphics[scale=\FIGSCALE, trim=35 0 0 0]{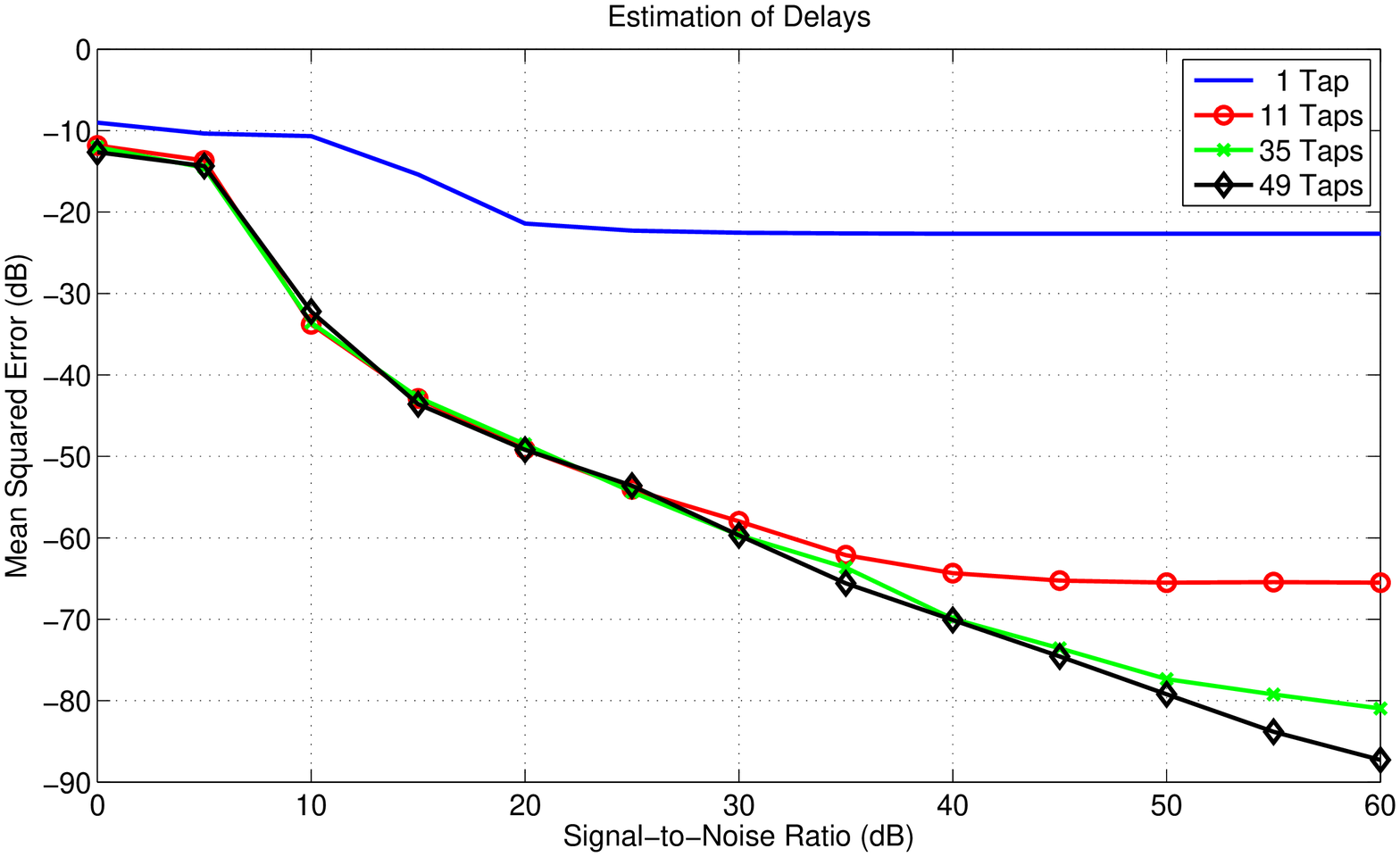}}%
\hfill%
\subfigure[]{\includegraphics[scale=\FIGSCALE, trim=35 0 0 0]{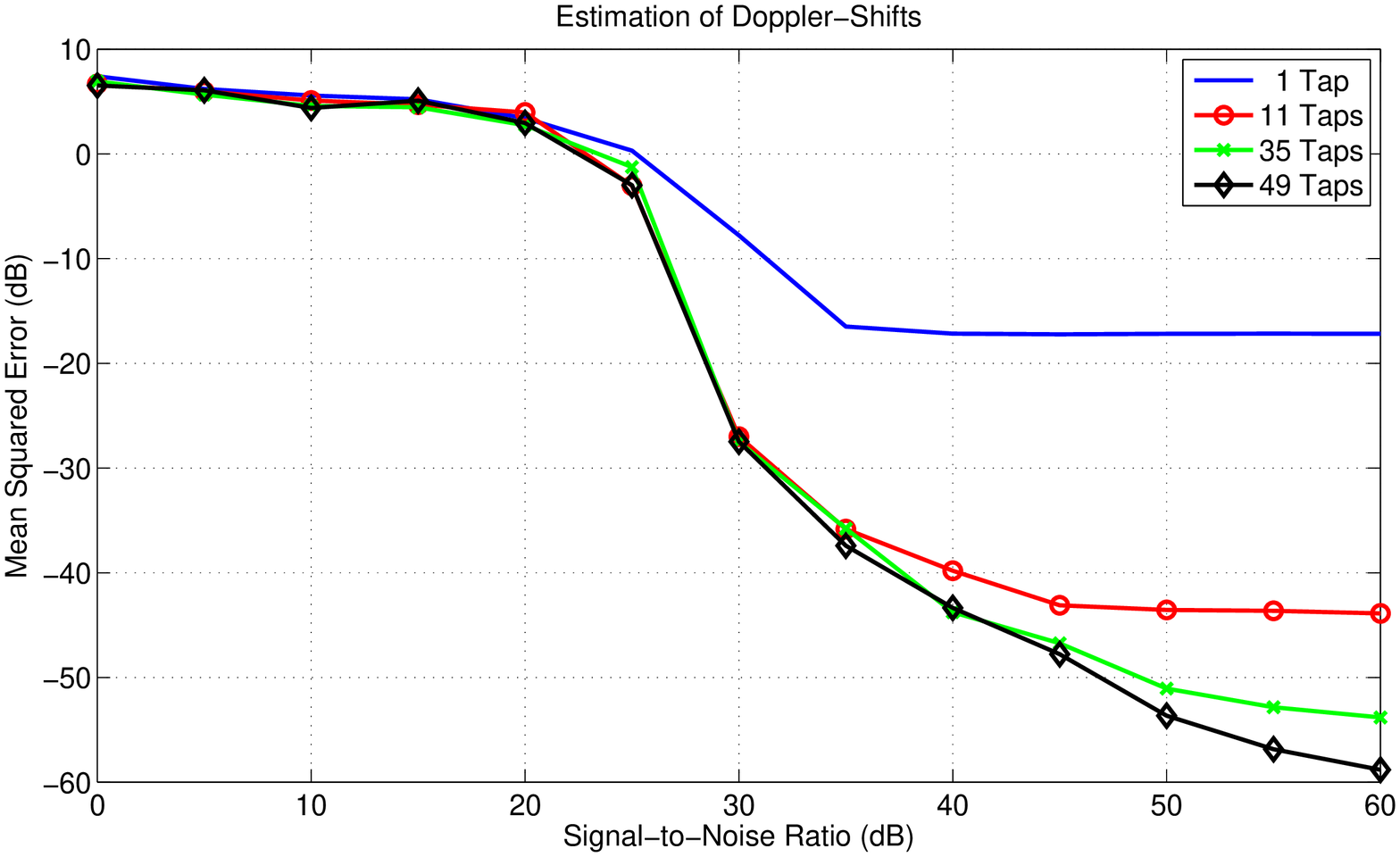}}%
\caption{Mean-squared error of the estimated delays and Doppler-shifts as a function of the signal-to-noise ratio for various lengths of the impulse responses of the filters.}%
\label{fig:VarDigLength}%
\end{figure}

Recall from Section~\ref{sec:id_sch} that our recovery method is composed of various digital-correction stages (see also Fig.~\ref{fig:samp_rec_scheme}). The filters used in these stages, which include $\{\phi_{\ell}[n]\}$ and $\{\psi_{\ell}[n]\}$, have infinite impulse responses in general so that their practical implementation requires truncation of their impulse responses. The truncated lengths of these filters also determine the (computational) delay and the computational load of the proposed procedure. Fig.~\ref{fig:VarDigLength} plots the MSE of the estimated delays (Fig.~\ref{fig:VarDigLength}(a)) and Doppler-shifts (Fig.~\ref{fig:VarDigLength}(b)) as a function of the SNR for various lengths of the impulse responses of the filters. There are two important insights that can be drawn from the results of this experiment. First, for a fixed length of the impulse responses, there is always some SNR beyond which the estimation error caused by the truncation of the impulse responses becomes more dominant than the error caused by the additive noise (as evident by the error floors in Fig.~\ref{fig:VarDigLength}). Second, and perhaps most importantly, filters with $49$ taps seem to provide good estimation accuracy up to an SNR of $60$ dB, whereas filters with even just $35$ taps yield good estimates at SNRs below $50$ dB.

\subsubsection{Effects of Finite Number of Samples}
\begin{figure}
\centering%
\subfigure[]{\includegraphics[scale=\FIGSCALE, trim=35 0 0 0]{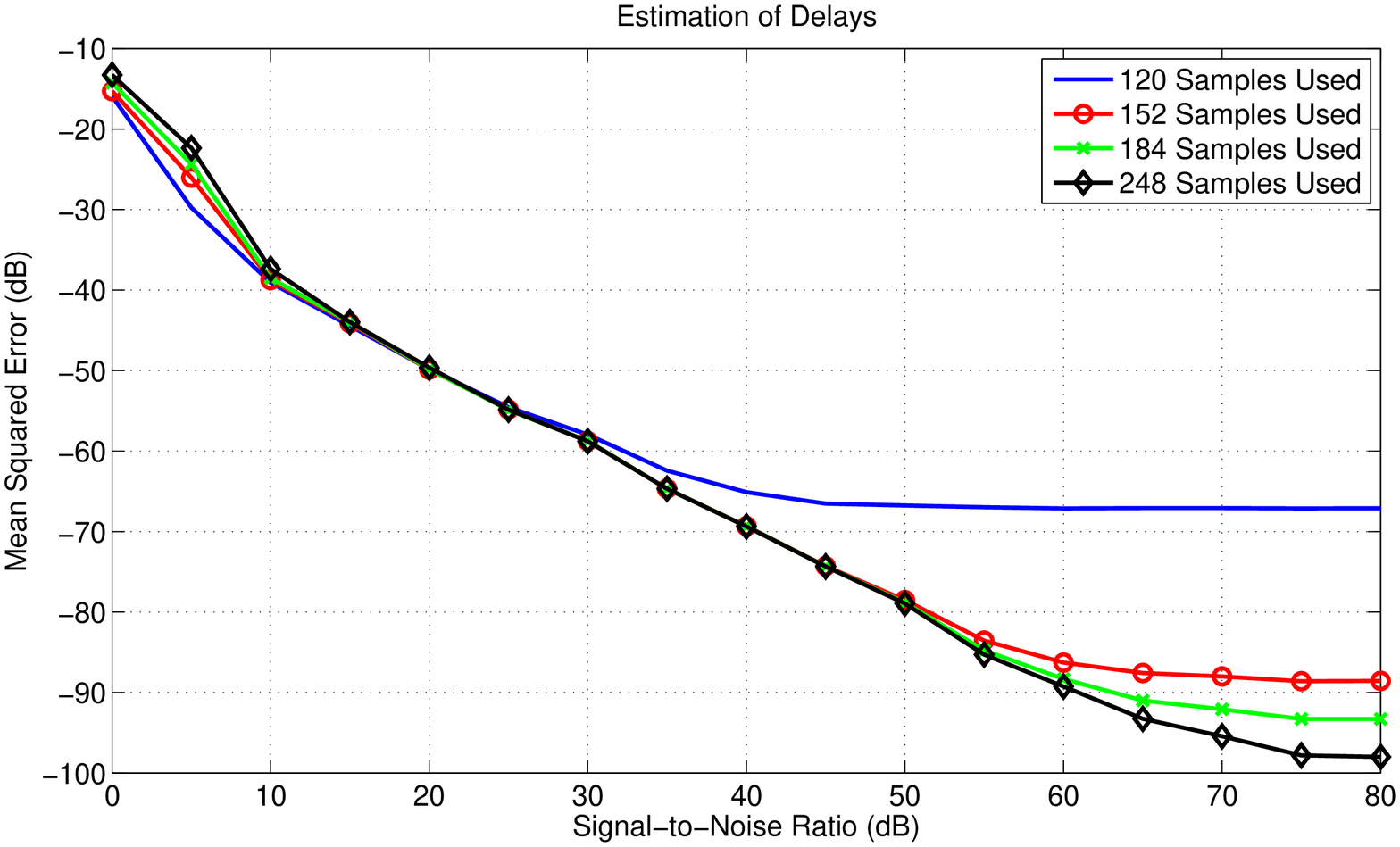}}%
\hfill%
\subfigure[]{\includegraphics[scale=\FIGSCALE, trim=35 0 0 0]{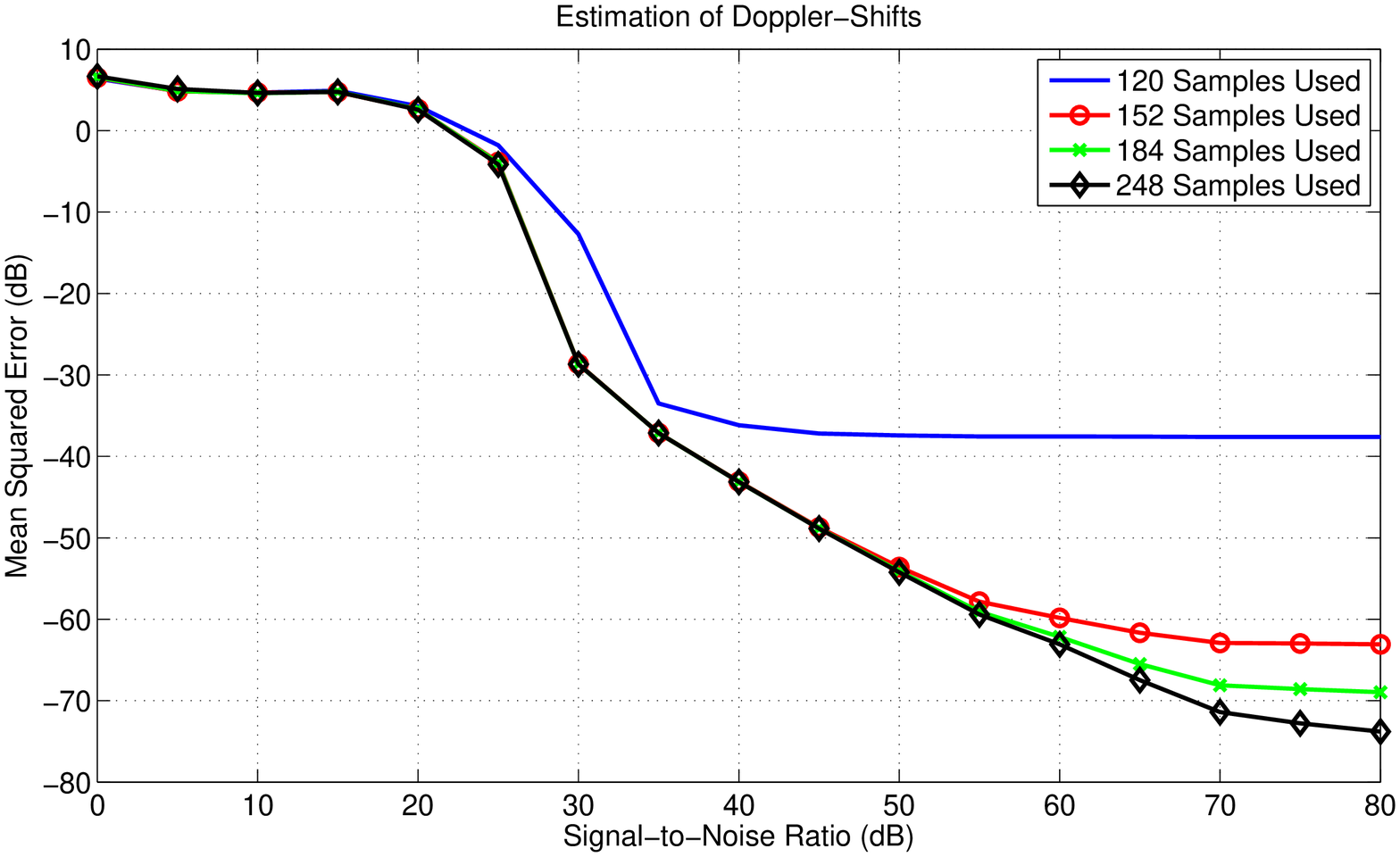}}%
\caption{Mean-squared error of the estimated delays and Doppler-shifts as a function of the signal-to-noise ratio for different numbers of samples collected at the output of the sampling filter (corresponding to an ideal low-pass filter).}%
\label{fig:VarSamples_sinc}%
\end{figure}

The sampling filter used at the front-end in Fig.~\ref{fig:samp_rec_scheme} is bandlimited in nature and, therefore, has infinite support in the time domain. Consequently, our sampling method theoretically requires collecting an infinite number of samples at the back-end of this filter. The next numerical experiment examines the effect of collecting a finite number of samples on the estimation performance. The results are reported in Fig.~\ref{fig:VarSamples_sinc}, which depicts the MSE of the estimated delays (Fig.~\ref{fig:VarSamples_sinc}(a)) and Doppler-shifts (Fig.~\ref{fig:VarSamples_sinc}(b)) as a function of SNR for different numbers of samples collected at the output of the sampling filter (corresponding to an ideal LPF). As in the case of truncation of digital-correction filter banks, there is always some SNR for every fixed number of samples beyond which the estimation error caused by the finite number of samples becomes more dominant than the error due to additive noise. Equally importantly, however, note that $x(t)$ in these experiments corresponds to a train of $N = 30$ prototype pulses. Therefore, under the assumption of $p = 4$ samples per pulse period $T$, it is clear that we require at least $N \cdot p = 120$ samples in total to represent just the input signal $x(t)$. On the other hand, Fig.~\ref{fig:VarSamples_sinc} shows that collecting $248$ samples, which is roughly twice the minimum number of samples required, provides good (delay and Doppler) estimation accuracy for SNRs up to $70$ dB.

\begin{figure}
\begin{center}
{\includegraphics[scale=0.355, trim=35 15 0 0]{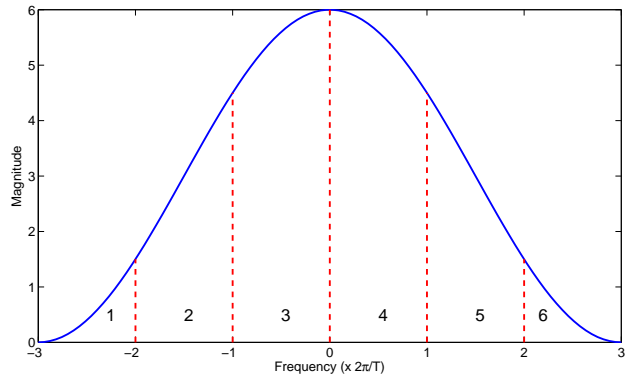}}
\caption{\label{fig:Rcos_freq}Frequency response of a raised-cosine filter with roll-off factor $1$.}
\end{center}
\end{figure}

Finally, it is worth noting here that making use of an ideal LPF as the sampling filter requires collecting relatively more samples at the filter back-end due to the slowly-decaying nature of the sinc kernel. Therefore, in order to reduce the number of samples required at the back-end of the sampling filter for reasonable estimation accuracy, we can instead make use of sampling filters whose (time-domain) kernels decay faster than the sinc kernel. One such possible choice is a raised-cosine filter with roll-off factor equal to $1$, whose frequency response is given by $S(\omega) = \frac{p}{2T}\left(1+\text{cos}(\frac{T}{p}\omega)\right)$ when $\omega \in \mathcal{F}$ and $S(\omega)=0$ when $\omega \not\in \mathcal{F}$. It is a well-known fact (and can be easily checked) that this filter decays faster in the time domain than the sinc kernel. However, the main issue here is that raised-cosine filter does not satisfy Condition~\ref{cond:filter_cond} in Section~\ref{sec:id_sch}, since its frequency response is not bounded away from zero at the ends of the spectral band $\mathcal{F}$ (see, e.g., Fig.~\ref{fig:Rcos_freq}).

\begin{figure}
\centering%
\subfigure[]{\includegraphics[scale=\FIGSCALE, trim=35 0 0 0]{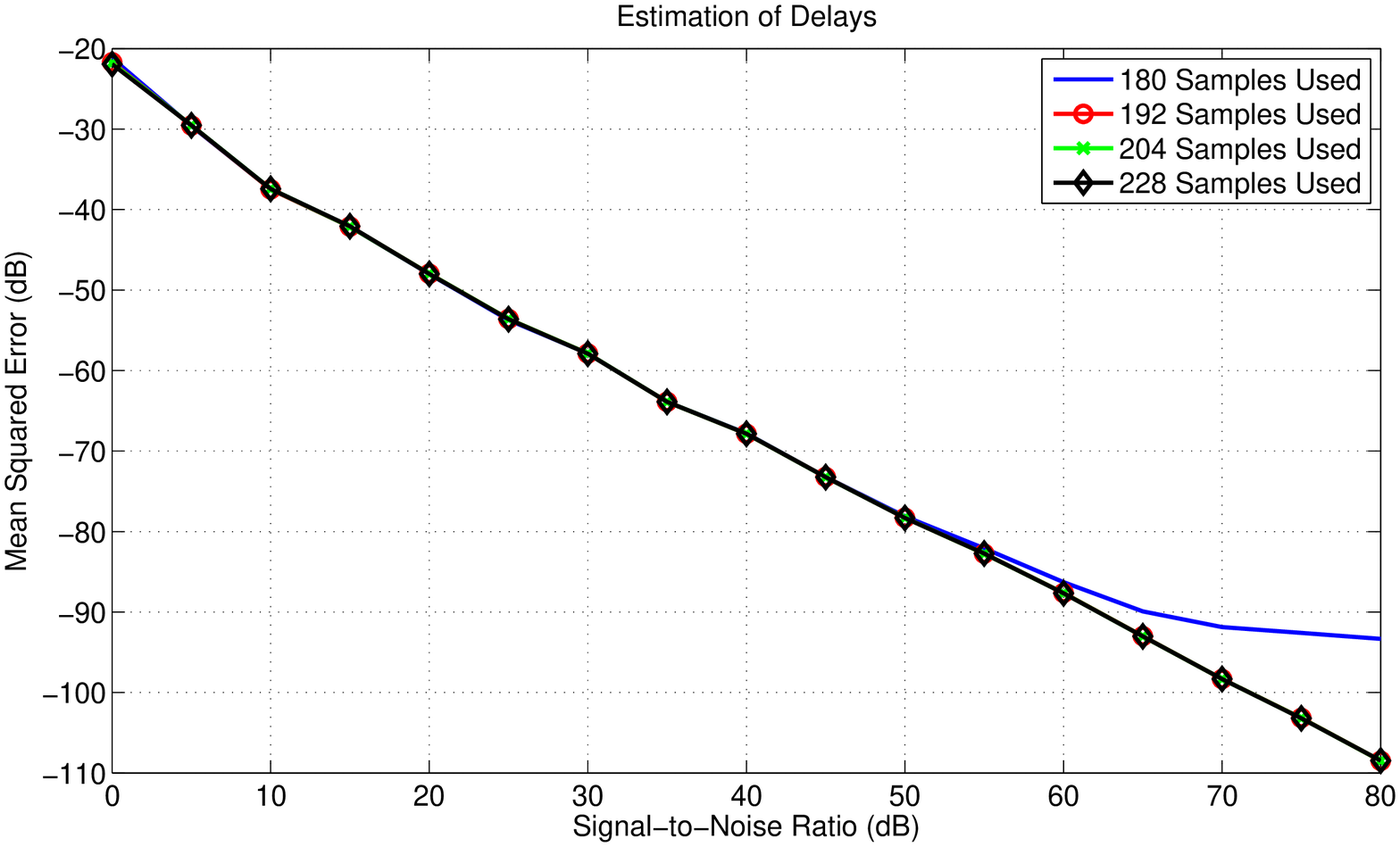}}%
\hfill%
\subfigure[]{\includegraphics[scale=\FIGSCALE, trim=35 0 0 0]{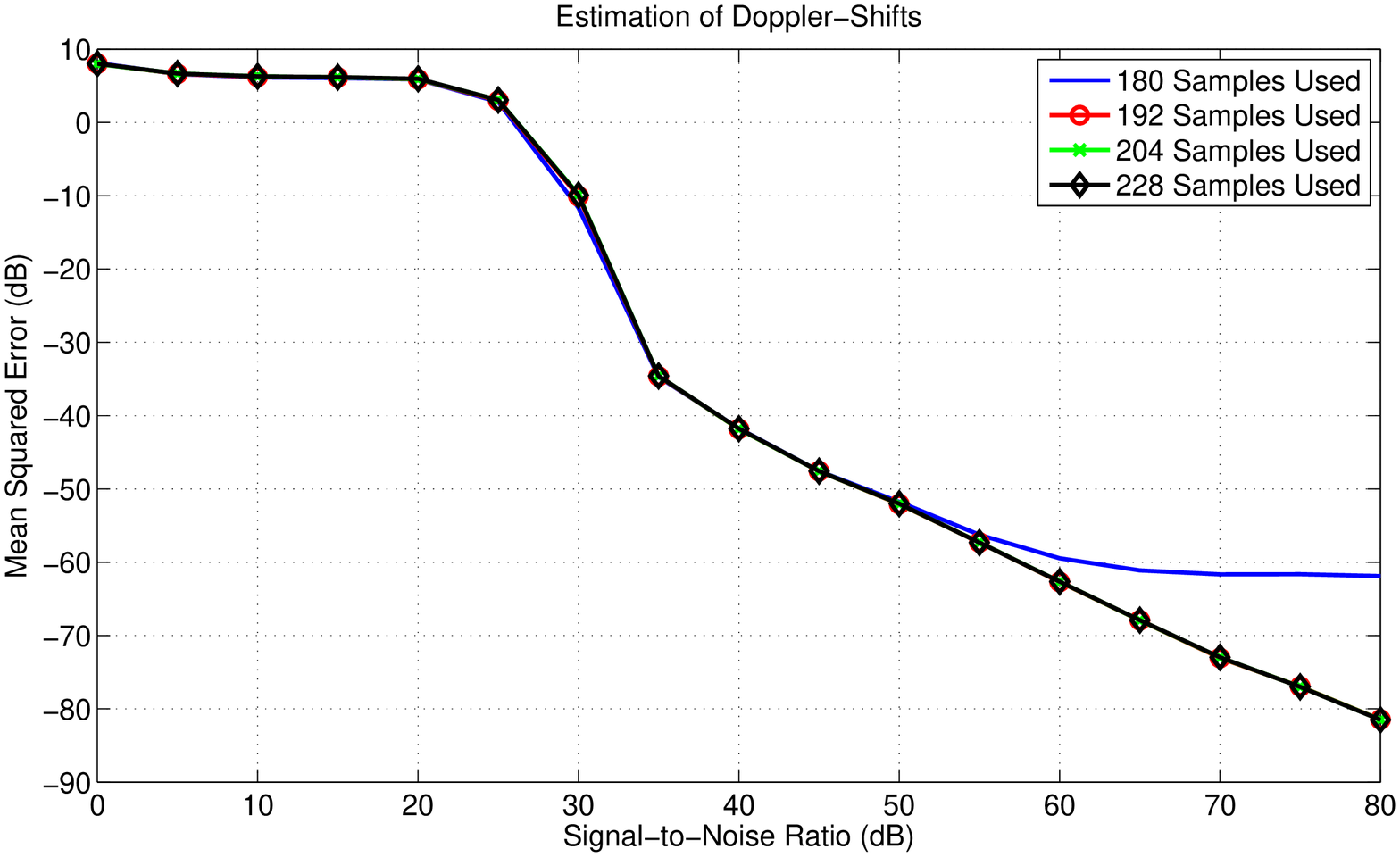}}%
\caption{Mean-squared error of the estimated delays and Doppler-shifts as a function of the signal-to-noise ratio for different numbers of samples collected at the output of a raised-cosine sampling filter with roll-off factor $1$.}%
\label{fig:VarSamples_rcos}%
\end{figure}

However, we now show that this problem can be overcome by slightly increasing the sampling rate and the bandwidth requirement stated in Section~\ref{sec:suff_cond}. Specifically, note that Proposition~\ref{pro:unique} requires that the parameter $p$, which controls the minimal bandwidth of $x(t)$ and the sampling rate of our proposed procedure, satisfies $p \geq 4$ under the current simulation setup (since $K_{\tau}=2$). We now instead choose $p = 6$ and argue that raised-cosine filter can be successfully used under this choice of $p$. To this end, recall from Section~\ref{sec:id_sch} that the function of the digital-correction filters $\psi_1[n]$ and $\psi_6[n]$ is to invert the frequency response of the sampling kernel corresponding to the frequency bands denoted by $1$ and $6$ in Fig.~\ref{fig:Rcos_freq}, respectively (under the assumption that the pulse $g(t)$ has a flat frequency response). In the case of a raised-cosine filter, however, we cannot compensate for the non-flat nature of these two bands since they are not bounded away from zero. Nevertheless, because of the fact that we are using $p = 6$, we can simply disregard channels $1$ and $6$ after the first digital-correction stage and work with the rest of the four channels ($2$-$4$) only. We make use of this insight to repeat the last numerical experiment using a raised-cosine filter and report the results in Fig.~\ref{fig:VarSamples_rcos}. It is easy to see from Fig.~\ref{fig:VarSamples_rcos} that, despite increasing $p$ to $6$, raised-cosine filter performs better than an ideal LPF using fewer samples.

\subsubsection{Effects of the Probing Sequence}

\begin{figure}
\centering%
\subfigure[]{\includegraphics[scale=\FIGSCALE, trim=35 0 0 0]{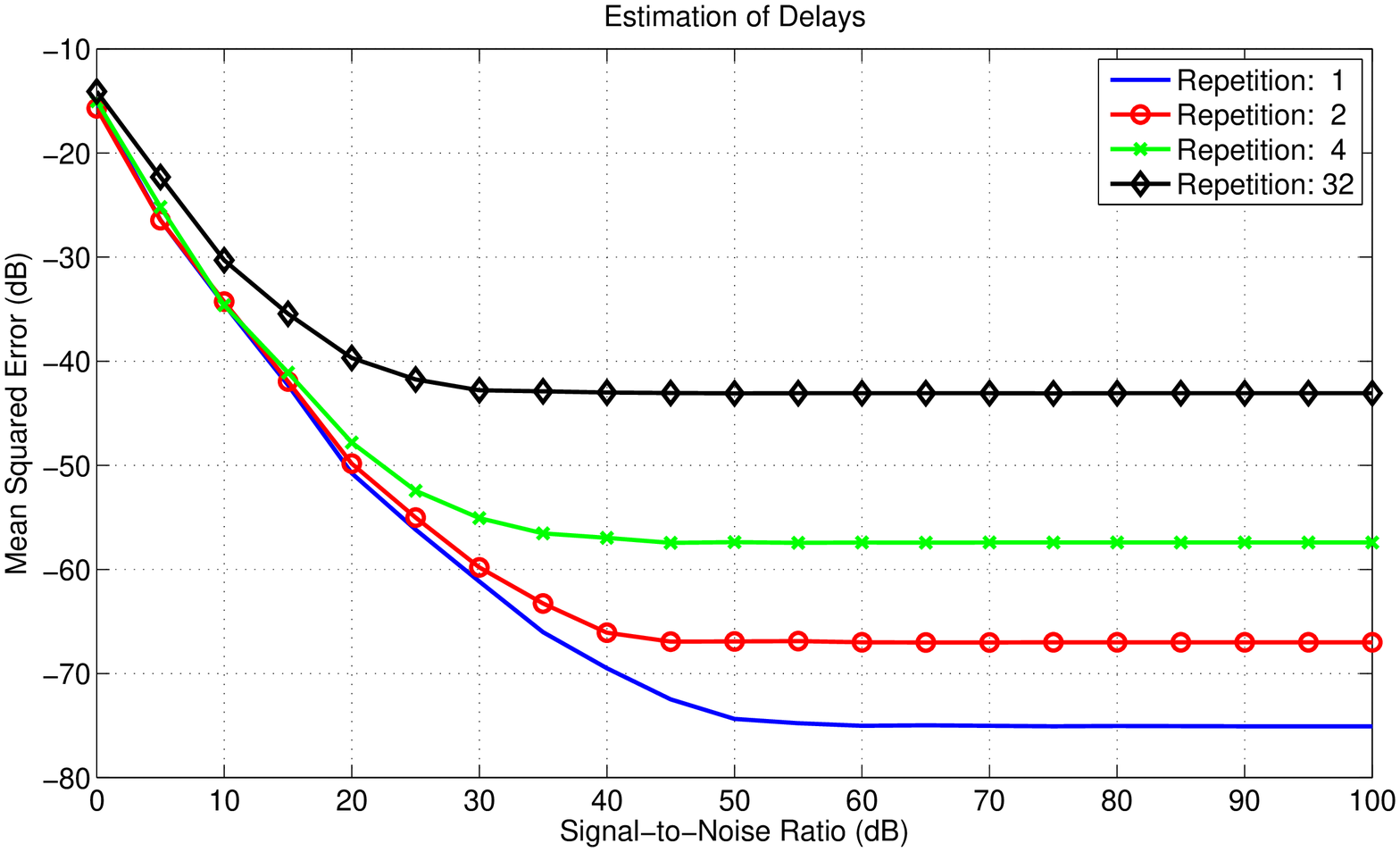}}%
\hfill%
\subfigure[]{\includegraphics[scale=\FIGSCALE, trim=35 0 0 0]{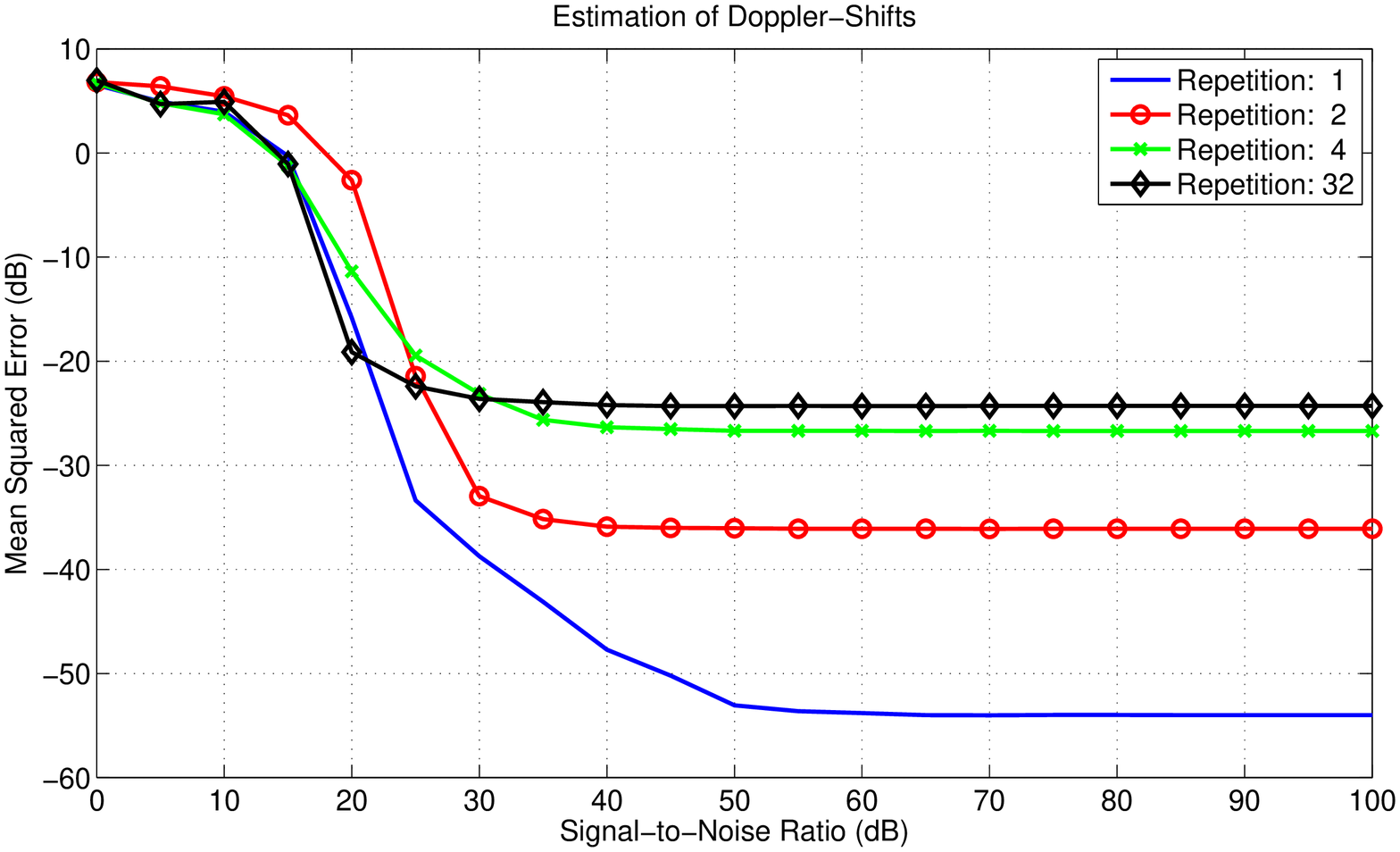}}%
\caption{Mean-squared error of the estimated delays and Doppler-shifts as a function of the signal-to-noise ratio for various probing sequences.}%
\label{fig:sequence}%
\end{figure}

Theorem~\ref{thm:summ_results} in Section~\ref{sec:suff_cond} stipulates that the choice of the probing sequence $\{x_n\}$ has no impact on the noiseless performance of the proposed recovery procedure as long as $|x_n| > 0$ for every $n=0,\dots,N-1$. However, it is quite expected that $\{x_n\}$ will have an effect on the performance in the presence of noise and implementation issues related to truncated digital filters and use of finite number of samples. The next experiment examines this effect for four different choices of binary probing sequences of length $N = 32$ that periodically alternate between $+1$ and $-1$ every $r$ entries. The results are reported in Fig.~\ref{fig:sequence}, which depicts the MSE of the estimated delays (Fig.~\ref{fig:sequence}(a)) and Doppler-shifts (Fig.~\ref{fig:sequence}(b)) as a function of the SNR for probing sequences with $r = 1, 2, 4,$ and $32$. We can draw two immediate conclusions from observing the results of this experiment. First, faster alternating probing sequences (in other words, sequences with higher frequency content) appear to provide better resilience against the truncation of digital filters and the use of finite number of samples. Second, the effect of the choice of probing sequence is less pronounced at low SNRs, since the error due to noise at low SNRs dominates the errors caused by other implementation imperfections.

\subsubsection{Effects of Model-Order Mismatch}
Our final numerical experiment studies the situation where the conditions of Theorem~\ref{thm:summ_results} do not exactly hold. To this end, we simulate identification of a parametric ULS with $K_{\tau}=4$ delays. For the first $3$ delays we take $K_{\nu,i}=2, \ i=1,2,3$, whereas we choose $K_{\nu,4}=8$ for the last delay. Finally, we take the prototype pulse $g(t)$ as at the start of this section (with bandwidth $\mathcal{W}=\frac{8\pi}{T}$), but we use a probing sequence $\{x_n\}$ corresponding to a random $(\pm 1)$ sequence with $N = 8$. Clearly, this does not satisfy the conditions of Theorem~\ref{thm:summ_results} because of the large number of Doppler-shifts associated with the last delay $(K_{\nu,4}=8)$.

\begin{figure}
\begin{center}
\includegraphics[scale=0.355, trim=35 15 0 0]{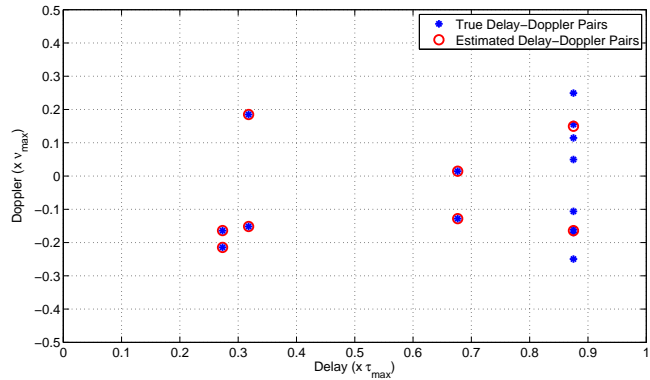}
\caption{\label{fig:Mismodel}Effects of model-order mismatch on the performance of the proposed recovery procedure corresponding to $\cH$ with $K=14$ delay--Doppler pairs.}
\end{center}
\end{figure}

The results of this numerical experiment are reported in Fig.~\ref{fig:Mismodel}. It can be easily seen from the figure that, despite the fact that $x(t)$ does not satisfy the conditions of Theorem~\ref{thm:summ_results}, our algorithm successfully recovers the first three delays and the corresponding Doppler-shifts. In addition, the fourth delay is correctly recovered but (as expected) the Doppler-shifts associated with the last delay are not properly identified. Note that in addition to demonstrating the robustness of our procedure in the presence of model-order mismatch, this experiment also highlights the advantage of the sequential nature of our approach where we first recover the delays and then estimate the Doppler-shifts and attenuation factors associated with the recovered delays. The main advantage of this being that if the input signal does not satisfy $N \geq 2K_{\nu,i}$ for some $i$ then recovery fails only for the Doppler-shifts associated with the $i$th delay. Moreover, recovery of the $i$th delay itself does not suffer from the mismodeling and it will be recovered correctly as long as the bandwidth of $x(t)$ is not too small.

% simulation file - DopplerDelaySNR_LPF_MisModel.m

%***************
\section{Conclusion}\label{sec:conc}
In this paper, we revisited the problem of identification of parametric underspread linear systems that are completely described by a finite set of delays and Doppler-shifts. We established that sufficiently-underspread parametric linear systems are identifiable as long as the time--bandwidth product of the input signal is proportional to the square of the total number of delay--Doppler pairs. In addition, we concretely specified the nature of the input signal and the structure of a corresponding polynomial-time recovery procedure that enable identification of parametric underspread linear systems. Extensive simulation results confirm that---as long as the time--bandwidth product of the input signal satisfies the requisite conditions---the proposed recovery procedure is quite robust to noise and other implementation issues. This makes our algorithm extremely useful for application areas in which the system performance depends critically on the ability to resolve closely spaced delay--Doppler pairs. In particular, our proposed identification method can be used for super-resolution target detection using radar.

%***************
\bibliographystyle{IEEEtran}
\bibliography{dd_sptrans10}

\end{document}